\documentclass{jfm}
\usepackage{graphicx}
\usepackage{epstopdf, epsfig}
\usepackage{comment}

\usepackage{hyperref}
\usepackage{dirtytalk}
\hypersetup{
    colorlinks=true,
    linkcolor=blue,
    filecolor=blue,      
    urlcolor=blue,
    citecolor=blue,
    }
\usepackage{soul}

\usepackage{color}

\usepackage[normalem]{ulem}
\setulcolor{red}
\sethlcolor{yellow}

\usepackage{xargs}                      
\usepackage[pdftex,dvipsnames]{xcolor}  
\usepackage[colorinlistoftodos,prependcaption,textsize=tiny]{todonotes}

\newcommandx{\unsure}[2][1=]{\todo[linecolor=red,backgroundcolor=red!25,bordercolor=red,#1]{#2}}
\newcommandx{\change}[2][1=]{\todo[linecolor=blue,backgroundcolor=blue!25,bordercolor=blue,#1]{#2}}
\newcommandx{\info}[2][1=]{\todo[linecolor=OliveGreen,backgroundcolor=OliveGreen!25,bordercolor=OliveGreen,#1]{#2}}
\newcommandx{\improvement}[2][1=]{\todo[linecolor=Plum,backgroundcolor=Plum!25,bordercolor=Plum,#1]{#2}}
\newcommandx{\thiswillnotshow}[2][1=]{\todo[disable,#1]{#2}}

\marginparwidth=\dimexpr \marginparwidth - 0.5cm\relax

\shorttitle{Advanced schlieren analyses for acoustic instabilities}
\shortauthor{S. L. Stahl, C. Kumar and D. V. Gaitonde}

\title{Shear-Layer Perturbation Responses from Time-Resolved Schlieren Data}
\author{Spencer L. Stahl\aff{1}
  \corresp{\email{spencer.stahl.ctr@us.af.mil}},
  Chandan Kumar \aff{2}
 \and Datta V. Gaitonde\aff{2}}

\affiliation{\aff{1}Air Force Research Laboratory, Wright-Patterson AFB, Ohio, 45433
\aff{2}Mechanical and Aerospace Engineering, The Ohio State University, Columbus, OH 43210, USA}

\begin{document}

\maketitle

\begin{abstract}
A novel combination of physics-based and data-driven post-processing techniques is proposed to extract acoustic-related shear-layer perturbation responses directly from spatio-temporally resolved schlieren video. The physics-based component is derived from a momentum potential theory extension that extracts irrotational (acoustic and thermal) information from density gradients embedded in schlieren pixel intensities. For the unheated shear layer, the method spotlights acoustic structures and tones otherwise hidden. The filtered data is then subjected to a data-driven Dynamic Mode Decomposition Reduced Order Model (DMD-ROM), which provides the response to forced perturbations. This method applies a learned linear model to isolate and quantify growth rates of acoustic phenomena suited for efficient parametric studies. A shear-layer comprised of two streams at Mach 2.461 and 0.175, corresponding to a convective Mach number 0.88 and containing shocks, is adopted for illustration. The overall perturbation response is first obtained using an impulse forcing in the wall normal direction of the splitter plate, extending in both subsonic and supersonic streams. Subsequently, impulse and harmonic forcings are independently applied in a local pixel-by-pixel manner for a precise receptivity study. The acoustic response shows a convective wavepacket and an acoustic burst from the splitter plate. The interaction with the primary shock and associated wave dispersion emits a second, slower, acoustic wave. Harmonic forcing indicates higher frequency-dependent sensitivity in the supersonic stream, with the most sensitive location near the outer boundary layer region. Excitation here yields an order of magnitude larger acoustic response compared to disturbances in the subsonic stream. Some receptive forcing inputs do not generate significant acoustic waves, which may guide excitation with low noise impact.
\end{abstract}


\section{Introduction}

\subsection{Overview}

Responses of flowfields to specifically placed perturbations is a crucial step in better understanding its properties and in determining suitable flow control strategies.
As such, a number of linear and nonlinear techniques have been developed to obtain responses to different types of perturbations.
Most employ the steady laminar or the time-mean turbulent flow obtained from variants of the Navier-Stokes equations as the basic state.
The goal of the present work is to extract acoustic-related perturbation responses from time-resolved schlieren data from uncontrolled experiments, using a relatively high-speed shear layer as a test bed for demonstration.

The proposed post-facto perturbation of uncontrolled experimental measurements presents exciting opportunities to non-intrusively study instabilities, but with many challenges.
Despite continual improvements in obtaining higher resolution unsteady data~\citep{Settles2017}, measurements are generally incomplete, in that only a subset of the domain and a few variables can be measured for any realization of the turbulent flow.
For example, Particle Image Velocimetry (PIV) only measures velocity~\citep{willert1991digital, adrian2011particle}, but not density or pressure, while schlieren images measure only density gradients in specific flow directions.
Although pixel intensity scales may be referenced, post-processing of time-resolved schlieren data is often limited to visual detection of phenomena such as shocks and shear-layers, but not rigorous perturbation stability analysis.

The proposed method leverages two steps to obtain the forced response from turbulent schlieren data.
In the first step, discussed in Section~\ref{secn:doak}, momentum potential theory (MPT), a physics-informed Kovasznay-like decomposition~\citep{Doak1998_FSE} adapted to filter density gradients in the schlieren data~\citep{PRASAD22} is used to extract acoustic-related components.
The second step, Section~\ref{secn:DMD-ROM},  applies the Dynamic
Mode Decomposition Reduced Order Model (DMD-ROM) data-driven procedure of
\citet{Stahl_2024_DMD}, to efficiently obtain the response and energy growth of defined forcing.
Originally developed for application to fields obtained from Navier-Stokes-based solutions, here the method is extended to obtain the response directly from the filtered schlieren data.


To demonstrate the new method, we use a planar compressible shear-layer formed by two parallel streams initially separated by a splitter plate, leveraging recently published data~\citep{KEVIN20B}, summarized in Section~\ref{subsec:schlieren}.
The Mach numbers of the upper (subscript $1$) and lower ($2$) streams are $M_1=U_1/a_1=2.46$ and $M_2/a_2=0.175$, respectively, where $M$ is the Mach number, $U$ is the streamwise velocity and $a$ is the speed of sound.
The parameter that principally characterizes the dynamics of the shear-layer is the convective Mach number, $M_c = (U_1 - U_2)/(a_1 + a_2)$.
Among the different cases of~\citet{KEVIN20B}, the $M_c = 0.88$ case is considered because it displays an interesting combination of coherent structures, shocks and attendant compressibility effects, which influence the shear-layer instabilities.
For this data, the receptivity and acoustic responses of the flow are extracted by considering various forcing characteristics at multiple locations across both streams.



\subsection{Context of the proposed approach} \label{sec:context}

A review of stability approaches provides context for the present approach.
Stability analysis examines the growth of small perturbations on a basic state.
Depending on the specific flow, the initial linear perturbation growth may develop into intricate non-linear multi-scale chaotic phenomena, which
make it difficult to fully model all stability perspectives from a single mathematical framework \citep{MENTER15}.
As such, the assumptions and exclusion of nonlinearity in many analytical and numerical approaches often results in poor prediction of experimental observations, particularly for high Reynolds and Mach number flows with complex geometry~\citep{stetson1988nonlinear,lachowicz1996boundary, lei2012linear}.

Stability analysis has evolved substantially since the pioneering shear-layer studies of \citet{helmholtz68,kelvin71}.
Fundamental developments may be found in various reviews \citep{schmid2007nonmodal,Theofilis2011} featuring many advanced computational approaches.
These methods usually invoke the Linearized Navier-Stokes Equations (LNSE).
Popular methods include linear stability theory (LST)~\citep{mack1984boundary}, parabolized stability equations (PSE), and nonlinear PSE (NPSE)~\citep{herbert1997parabolized}, each of which considers  local spatial or temporal stability on projections of the flowfield.
For example, LST assumes parallel flow; PSE and NPSE are favored when nonparallel or nonlinear effects are important, at the expense of exponentially increased computational costs \citep{Juniper2014}.
Three-dimensional global analyses are also becoming increasingly common~\citep{johnson2010three, paredes2014advances}, but are considerably more expensive to compute.
A distinct yet related LNSE based approach is resolvent analysis \citep{TRE99, JOVA05, SHARMA10}, which pursues input-output forcing-response pairs and sensitivities for optimal instability growth~\citep{taira2017turbulent}.

The proposed method shares similar goals as these traditional stability analyses, but is unique in the sense that it uses schlieren diognostics, which is advantageous as well as scope-limiting.  
The advantage is that unlike LNSE-based methods which require all flow variables, the current approach operates with schlieren measurements, which only diagnose density gradients.
This necessitates an element of data-driven approaches; even so, as is demonstrated below, density gradients by themselves are of limited value from a perturbation propagation perspective since they can yield incoherent results. 

To address this, the schlieren data is first processed to filter Kovasznay-like acoustic components that facilitate clearer analysis (Section~\ref{secn:doak}).
When all variables are available from a scale-resolved simulation, MPT exactly decomposes ``momentum density'' into acoustic (irrotational-isentropic), thermal (irrotational-isobaric) and vortical (rotational) components, regardless of flow nonlinearity or fluctuation magnitude.
This decomposition has been used to study instabilities in jets~\citep{unnikrishnan2019interactions} and hypersonic transition~\citep{Unnikrishnan_Gaitonde_2021}.
For schlieren data, an MPT extension by~\citet{PRASAD22} extracts the irrotational component; in unheated flows of interest here, where thermal fluctuations are not significant, the resulting field represents the acoustic dynamics.
Filtering the schlieren data successfully extracts various aspects of shear layers not apparent from processing of the raw schlieren data, including Mach wave radiation, feedback mode shapes, screech signatures and coupling between adjacent jets~\citep{prasad2022examination}.

Although MPT-based filtering effectively isolates schlieren content representative of acoustic phenomena, it does not provide stability analysis.
For this, the second component of the proposed method uses a reduced model to obtain the transient forced response.
Data-driven modal decomposition of the unsteady fluctuations is a natural choice upon which to build the desired model.
The most common approach, Proper Orthogonal Decomposition (POD), yields modes of correlated spatial structures ranked by energy \citep{Berkooz1993,Sirovich1987}.
Spectral POD \citep{Lumley_POD,TOWNE18,Schmidt2020} and conditional space-time variants \citep{Schmidt2019} (SPOD and CST-POD, respectively) can also generate insights on instabilities, including signatures of absolute (tonal) and convective instabilities and have been used for various problems including shear-layers \citep{Stack2019,LU2019263} and related jets \citep{Berkooz1993,boree2003extended,Weightman2018,towne2018spectral,Karami2018a_spod}.
Recent examples for processing schlieren data to extract coherent structures in such flows may be found in \citet{Weightman2017,Liu2021_sijmoswit,BERRY17}.
 


Here we use Dynamic Mode Decomposition (DMD) as the basis; DMD takes a dynamical systems approach to form a ``best fit'', reduced linear operator that approximates non-linear dynamics based on Koopman theory \citep{Schmid2007,SCH11}.
As a modal decomposition technique, time-dependent datasets may be reduced by POD coordinates, enabling DMD eigen-modes analogous to those of the LNSE operator.
Connections between DMD modes obtained from statistically stationary data, SPOD, and resolvent modes have been identified by \citet{TOWNE18}, with further connections available when used in conjunction with CST-POD to isolate causal forcing mechanisms and ensuing shear-layer instabilities \citep{stahl_2023_CPOD_jcp}.
Other approaches have leveraged the connection between DMD and stability modes in combination with LNSE-generated data.
For example, the Mean Flow Perturbation (MFP) method obtains unstable tri-global modes that, unlike local LST, propagate linear disturbances on a basic state before subjecting them to DMD \citep{RAJ18}.

In the present work, DMD is used not for model reduction, but as a basis for the reduced order model (DMD-ROM), specifically by using the linear operator to evolve the dynamics forward in a reduced subspace, as discussed in Section~\ref{secn:DMD_ROM}.
The stability analysis is formulated by applying a forcing term, representing the imposed perturbation, to the DMD-ROM, and observing and post-processing the temporal response of the disturbed flow.
As demonstrated by \cite{Stahl_2024_DMD}, the approach constitutes a versatile and efficient framework, suitable for parametric studies of instability dynamics, the receptivity of different forcing parameters, and to guide flow control. 
In that work, harmonic pressure forcing using the DMD-ROM correctly predicted the effects of blowing-suction control of a supersonic impinging jet shear-layer.
For time-local convective instabilities, the DMD-ROM was used with CST-POD to identify the optimal forcing phase that diminished or amplified convective instabilities.
In the current work, we considerably extend the approach by combining MPT with the DMD-ROM to induce and isolate such instabilities directly from schlieren data using impulse and harmonic forcing.

\subsection{Compressible shear-layer instabilities}\label{secn:intro_SL}
\begin{figure}
  \centerline{\includegraphics[width=1\textwidth]{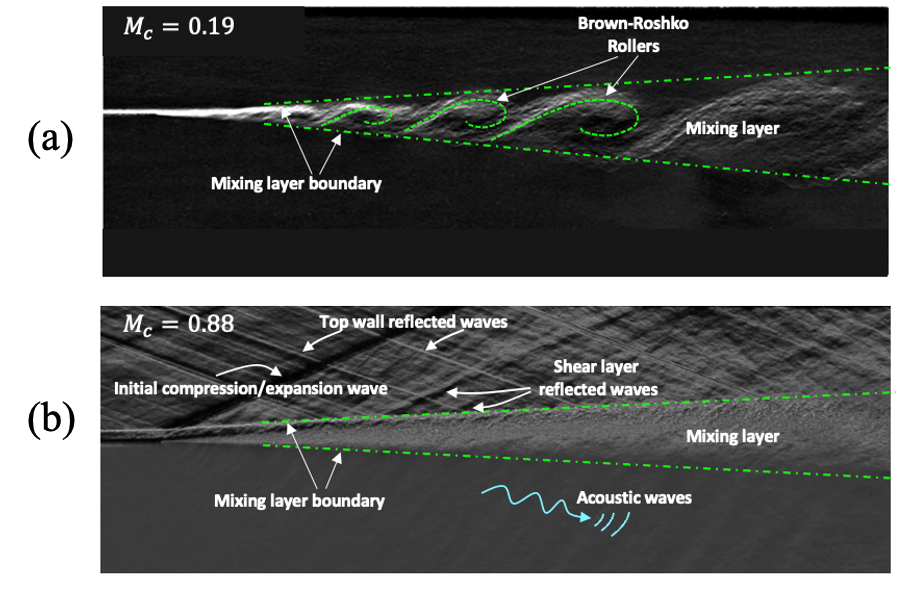}}
  \caption{Flow features of free shear-layers for (a) $M_c = 0.19$ (b) $M_c = 0.88$. Large Brown-Roshko roller structures at lower speeds are replaced by fine scale turbulent mixing and the presence of shocks at higher convective Mach numbers. 
  Schlieren images from~\cite{KEVIN20B} with labels added to highlight the features.}
\label{fig:schlieren_intro}
\end{figure}

Shear-layer instabilities are associated with inflection points in velocity gradients across adjacent fluid layers~\citep{browand1966experimental,michalke1972instability}.
Small perturbations from the inflow environment or near the splitter plate trailing edge excite instability waves that grow downstream.
At low speeds, these waves roll up into vortical structures~\citep{nitsche2006vortex};
subsequently, three-dimensional effects deform the structures through spanwise undulations, eventually resulting in a fully-developed turbulent region characterized by self-similar time-averaged velocity and Reynolds stress profiles~\citep{ launder1975progress, ROGERS94}.
The convective Mach number, $M_c$, is related to the speeds of the largest structures in the mixing-layer region, and also identifies compressibility effects related to the growth rates of instability waves~\citep{papaR1988}. 
Compressibility effects typically become significant when $M_c>0.5$ and increase rapidly with the Mach number of the supersonic stream, as shocks of increasing strength augment the nonlinear development of large-scale structures; these effects asymptote beyond $M_c>1$~\citep{PAPA88}.
Under such conditions, the generalized inflection point associated with density or temperature profile variations becomes important \citep{leib1991nonlinear}. 
The chosen $M_c = 0.88$ experiments of~\cite{KEVIN20B, KEVIN20,Dutton2019} thus provides a rich compressible dataset to test the new analysis method proposed in this work.
Schlieren images provide excellent visualization of compressible shear-layer instabilities.
For illustration, instantaneous snapshots from \citet{KEVIN20B} are referenced in Fig.~\ref{fig:schlieren_intro} at two convective Mach numbers, (a) $M_c=0.19$ and (b) $M_c=0.88$.
At lower $M_c$, the coherent structure roll-up is apparent in Fig.~\ref{fig:schlieren_intro}(a).
\citet{BROWN74} provide a comprehensive discussion of the shape, organization, and dynamics of such large-scale Brown-Roshko roller structures found in incompressible planar mixing layers. 
The initially round structures transition to flattened and elongated waves with increasing $M_c$.
\citet{ELLI95} similarly observed stretched shear-layer structures undergoing a pairing process at $M_c=0.51$, resembling Kelvin-Helmholtz type structures. 
However, as convective Mach number increased to $M_c=0.85$, no large-scale structures are observed as the mixing layer quickly transitions to turbulence.
This behavior is also evidenced in the relevant $M_c=0.88$ experiments of \cite{KEVIN20B}, and visualized in Fig.~\ref{fig:schlieren_intro}(b), where the mixing layer behind the shock rapidly transitions to fine-scale turbulence.
The increased compressibility enhances the anisotropic behavior of streamwise normal stresses, consistent with the observed stretching and breakdown of smaller structures.
Note in Fig.~\ref{fig:schlieren_intro}(b), that the turbulent mixing-layer appears noisy in schlieren images, obscuring the source of acoustic waves emanating into the lower stream; a phenomenon revealed by our filtering approach. 



Additional complexities are encountered at higher convective Mach number flows with shocks.
Returning to Fig.~\ref{fig:schlieren_intro}(b), the shock system is initiated at the splitter plate trailing edge and undergoes a series of expansions and reflections in the upper stream between the top wall and upper boundary of the mixing layer.
The strength of the compression depends on the pressure difference across the splitter plate, which is equalized through the shocks, weakening downstream.
\cite{KEVIN20B} examined the effect of the shock, reporting that the turbulent statistics of the mixing layer were augmented by increased Reynolds stresses.
An important byproduct of the shock interaction is emission of acoustic energy, which was not characterized by the schlieren or PIV results but is relevant for shear-layer instabilities, for example, in the context of jet noise.
Elsewhere, in simple and complex shear-layer configurations~\citep{berry2017application},
unsteady shock interactions arise with prominent acoustic waves  from shedding at the splitter plate again suggesting the underlying importance of acoustics processes.

The background disturbance environment plays an important role in influencing shear-layer instabilities, and is therefore relevant to the present forced-response analysis.
Tunnel noise is unavoidable, and the ``natural'' instability mechanisms from stochastic disturbances are well-documented \citep{hussain_1987}. 
\cite{SAM18} provide a recent overview of flow structures in high-speed shear-layers, noting that naturally generated perturbations mostly belong to a broadband spectrum. 
The proximity of the excitation frequency to the most-amplified instability frequency can lead to a lock-on control situation, suppressing the natural pairing process of large-scale structures.
For mixing layers, a numerical study indicated that two-dimensional and weakly oblique structures form downstream of the splitter plate \citep{SAND10}.
However, these structures lack the high coherence observed in experiments \citep{PAPA94}, pointing to the existence of forcing noise components that are not accounted for in the simulations. 

Other experiments investigate shear-layer stability by intentionally forcing the flow.  
For example, \cite{DEN92} used a glow discharge forcing technique to excite the shear-layer. 
At higher Mach numbers, additional oblique instability waves dominate, which are absent in the unforced experiments. 
\cite{MAR96} similarly conducted detailed measurements of high-speed Mach~3 and~4 streams with a low-speed Mach~1.2 stream, using a glow discharge to excite planar and oblique waves.
Planar excitation remained planar, but oblique excitation changed the observations significantly, with the emergence of different wave propagation angles than initially used for excitation.
Increasing the convective Mach number yielded slightly more oblique instability waves, consistent with the results of \citet{DEN92}.
Forcing insights like these have lead to promising control techniques for supersonic shear-layers \citep{DOSHI22,Oka2024}. 

The aforementioned research motivates the desire to understand shear-layer receptivity and its response to different forcing inputs such as frequency, location, and orientation.
For experiments that intentionally force the shear-layer \citep{DEN92,MAR96,papaR1988,SAM18}, these questions can be directly pursued through parametric studies. 
However, many experiments examine the unforced system and only elicit the natural instabilities excited by wind tunnel noise, leaving uncertain the response to specific forcing.
While PIV measurements may be used in a post-facto analysis to test forced stability or noise generation mechanisms under these circumstances \citep{sinha2012impulse,Ghasemi2018}, no equivalent analysis using schlieren data has yet been conducted, which we address in this paper.

\begin{figure}
   \centerline{\includegraphics[width=1\textwidth]{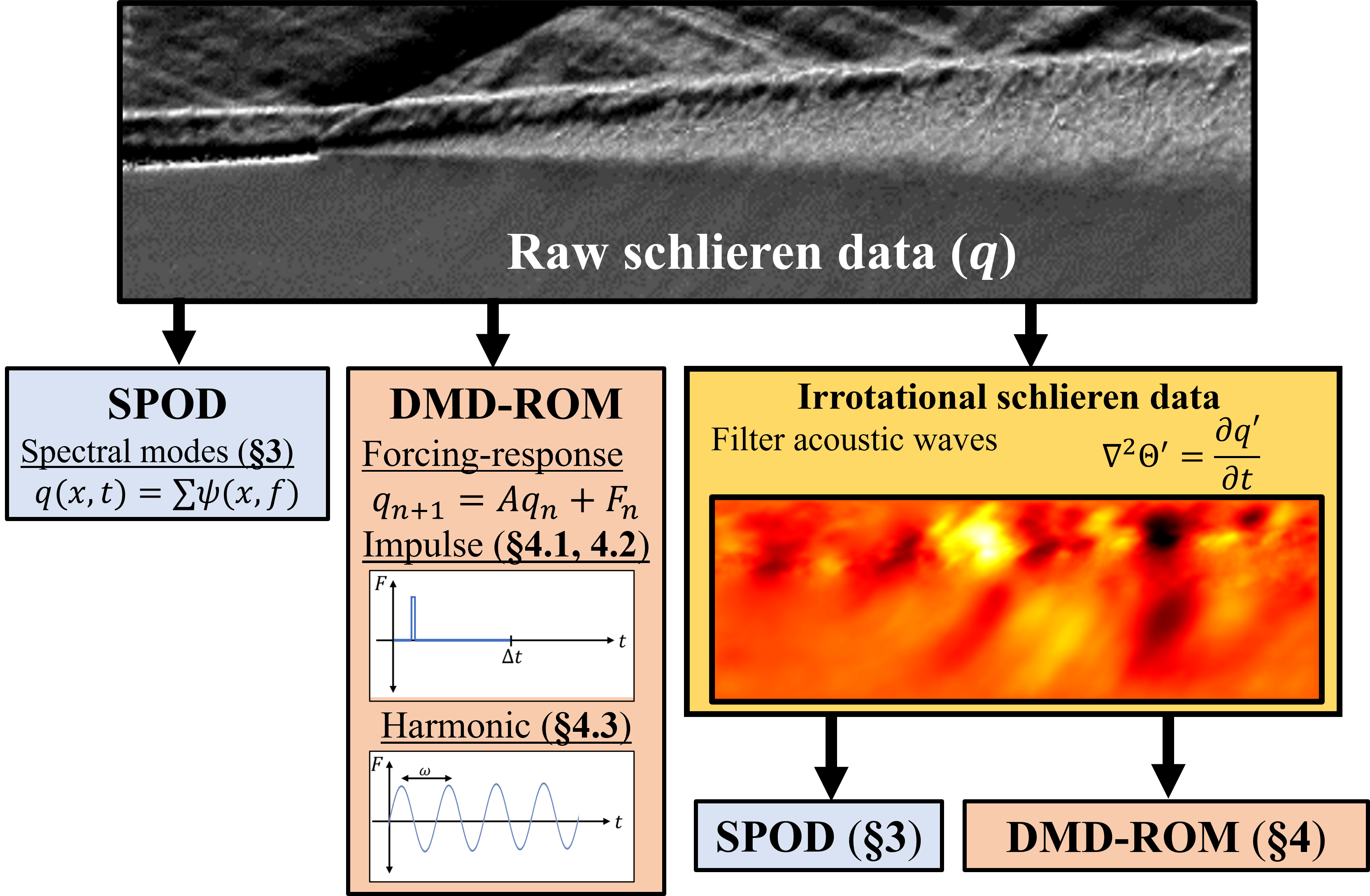}}
  \caption{Outline of schlieren post-processing analyses comparing the raw and filtered schlieren data. SPOD analysis is conducted in Secn.~\ref{secn:SPOD} and a DMD-ROM analysis is presented for impulse and harmonic forcings in Secn.~\ref{secn:DMD_ROM}.}
\label{fig:overview}
\end{figure}

The paper is organized as follows.  
For reference, the overall approach of the study is outlined in Fig.~\ref{fig:overview}.
Each of the two steps in the procedure is described in Section~\ref{secn:method}, including the filtering process in~\ref{secn:doak} and the DMD-ROM forcing framework~\ref{secn:DMD-ROM}.
The features of the raw schlieren data are outlined in~\ref{subsec:schlieren}.  
An ancillary SPOD analysis in Section~\ref{secn:SPOD} demonstrates the advantages of the filtered versus the raw schlieren data.
In Section~\ref{secn:DMD_ROM}, the DMD-ROM is implemented to examine the forced response of the system. 
An impulse forcing is first applied to isolate the shear-layer convective instability, followed by a harmonic forcing analysis to examine the asymptotic dynamics.
Parametric receptivity studies test the frequencies and locations in the upper and lower streams that most amplify the acoustic response.
Finally, conclusions are summarized in Section~\ref{secn:conclusion}.


\section{Methodology}\label{secn:method}

 
\subsection{Irrotational filtering}
\label{secn:doak}

The MPT procedure used to extract the irrotational component from a flow is first summarized in its original form \citep{DOAK1989}, suitable for use when complete data is available as from scale-resolved simulations (e.g., \citet{UNNI16}) before describing its adaptation to unsteady schlieren data \citep{PRASAD22}. 
MPT applies the Helmholtz decomposition to the ``momentum-density''
vector field, $\rho\textbf{u}$ to obtain  mean and fluctuating rotational (solenoidal) components, $\overline{\textbf{B}}$ and  $\textbf{B}'$, and the irrotational fluctuating component, $\nabla\psi'$:
\begin{equation}
    \rho \textbf{u}=\overline{\textbf{B}} + \textbf{B}^{'} - \nabla\psi^{'}.
    \label{equation:hlmhtz}  
\end{equation}
The solenoidal momentum is, by definition, divergence free,
\begin{equation}
\nabla\cdot\overline{\textbf{B}}=0~,~~~~~\nabla\cdot\textbf{B}^{'}=0,
\end{equation}
while the irrotational potential is curl free,
\begin{equation}
\nabla \times \psi'=0.
\end{equation}
The scalar potential $\psi$ comprises mean $\overline{\psi}$, and fluctuating components $\psi^{'}$. 
Given a statistically stationary flow, the mean scalar potential gradient ($\nabla\overline{\psi}$) is zero. 
The scalar potential is calculated by substituting Eqn.~\ref{equation:hlmhtz} into the conservation of mass equation,
\begin{equation}
    \frac{\partial \rho}{\partial t} + \nabla \cdot ( \rho \textbf{u})=0,
\end{equation}
resulting in the Poisson equation, 
\begin{equation}\label{eqn_poi_ir}
    \nabla^{2}\psi^{'}=\frac{\partial \rho^{'}}{\partial t}.
\end{equation}
The fluctuating scalar potential is comprised of acoustic (isentropic) $\psi^{'}_{A}$, and thermal (isobaric) $\psi^{'}_{T}$ components,
\begin{equation}
    \psi^{'} = \psi^{'}_{A} + \psi^{'}_{T},
 \end{equation}
which are uniquely expressed as, 
 \begin{equation}   \label{eqn_psi_A}
 \nabla^{2}\psi^{'}_{A}=\frac{1}{c^{2}}\frac{\partial P^{'}}{\partial t}~,~~~~~~~~~~~~~ \nabla^{2}\psi^{'}_{T}=\frac{\partial \rho}{\partial S}\frac{\partial S^{'}}{\partial t}.
\end{equation}
Here $c$ is the instantaneous local speed of sound, $P$ is the thermodynamic pressure, and $S$ is the entropy. 

Schlieren data provides density gradients; the form of Eqn.~\ref{eqn_poi_ir} where the time-derivative of density appears as the forcing term can thus be leveraged to isolate the irrotational component.
As discussed in \citet{PRASAD22}, since the raw schlieren pixel intensity, $q$, is proportional to the streamwise density gradient,
\begin{equation}
    q \propto \frac{\partial \rho}{\partial x},
\end{equation}
MPT may be used to solve for the appropriate derivative (streamwise in this case) of the irrotational scalar potential $\psi'$ in the Poisson relation of Eqn.~\ref{eqn_poi_ir}:
\begin{equation}
\nabla^{2}(\frac{\partial \psi'}{\partial x})=\frac{\partial}{\partial t}(\frac{\partial \rho}{\partial x}).
\end{equation}
A quantity proportional to the irrotational momentum in the streamwise direction, $\Theta'$, can be directly solved for by substituting the time derivative of pixel intensity.
 \begin{equation}\label{eqn_poi_ir_schl}
 \nabla^{2}\Theta^{'}=\frac{\partial q^{'}}{\partial t}.
 \end{equation}
This irrotational filtered quantity, $\Theta'$, has contributions from both acoustic and thermal components.
Previous numerical studies \citep{UNNI18, PRASAD20, PRASAD21} have shown the thermal component exhibits coherent structures similar to its acoustic counterpart but is non-radiating, even for heated flows.
For unheated flows, changes in the irrotational component are dominated by those in the acoustic component.
Outside of the shear-layer region, radiated density fluctuations are purely acoustic and thus captured in $\Theta'$, as shown by \cite{PRASAD22}.


The solution of a Poisson equation requires suitable boundary conditions.
Several prior efforts \citep{UNNI18, PRASAD20, PRASAD21,stahl2021_aviation} have explored the sensitivity of the results to boundary condition specifics.
When the boundaries are far from regions of rapid gradients, suitable conditions may be obtained by integrating the momentum vector to obtain the irrotational boundary conditions.
This solution has been determined to be relatively insensitive to boundary conditions when these are applied far from the regions of interest, since the potential function is dominated by source term fluctuations.
For schlieren data on the other hand, the $\rho \boldsymbol{u}$ is not available and the boundaries may not lie in regions of low vorticity.
Following \citet{PRASAD22}, an effective procedure is to use a Dirichlet boundary condition of $\Theta'=0$ with a hyperbolic tangent sponge zone extending eight pixels from the border. 
The thin splitter plate is not treated as a boundary condition due to the limited pixel resolution and sponge zone which would damp the boundary-layer fluctuations.
While this decreases spatial resolution across the splitter plate region, the resulting wavepacket signatures are nonetheless dominated by the Poisson equation source term (i.e., pixel fluctuations) and represent the dynamics on the upper and lower sides, with no residual effect further downstream.
An alternative approach to process schlieren data using MPT in the frequency domain has been discussed by \citet{padilla2024eduction}.

\subsection{DMD-ROM forcing and response}\label{secn:DMD-ROM}

\begin{figure}
\centering
\includegraphics[width=1\textwidth]{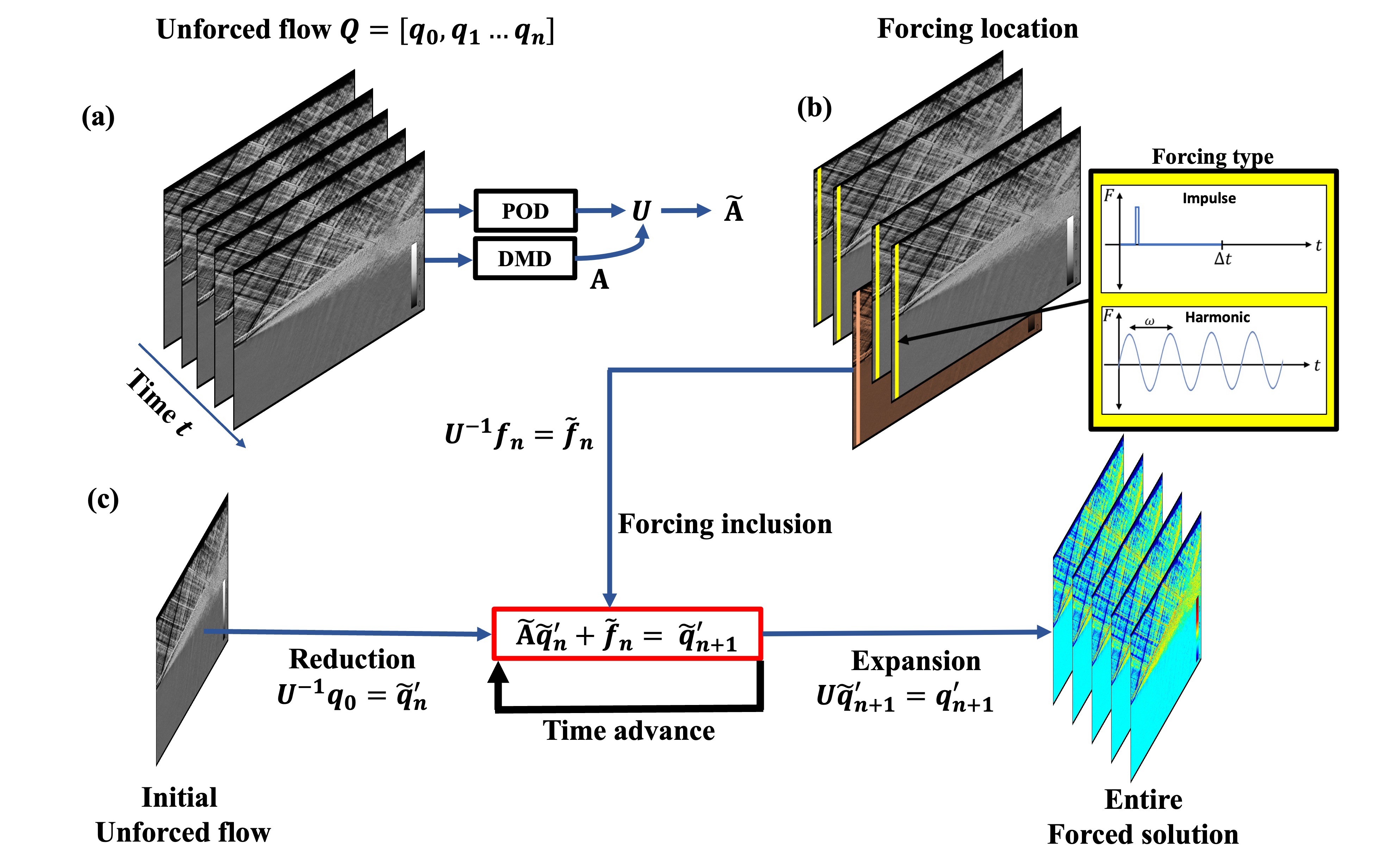}
\caption{DMD-ROM forcing methodology. (a) Input flow snapshots provide the DMD linear operator $\boldsymbol{\tilde{A}}$ and POD modes $\textbf{U}$ for snapshot reduction. (b) User defined forcing snapshots. (c) DMD-ROM time marching scheme featuring the reduction of the initial condition, time advancement with forcing, and the final expansion of the forced solution.}
\label{fig:DMD_method}
\end{figure}

The shear-layer stability and receptivity study is enabled by applying a forcing term to a DMD-ROM, as outlined in the framework of \citet{Stahl_2024_DMD}, where it was employed with the unsteady pressure field obtained from a scale-resolving simulation.
An overview of the method is shown in Fig.~\ref{fig:DMD_method}.
The DMD-ROM is derived from the (a) snapshot data matrix of either the raw or MPT-filtered schlieren data, $\textbf{Q}=[q_1,...,q_N]$, where there are $N$ individual snapshots, $q_n$, of size $M$ pixels.
DMD ultimately seeks a linear operator, $\textbf{A}$, that marches the system from one snapshot, $q_n$, to the next, $q_{n+1}$:
\begin{equation}\label{eqn:dmd_snap_control}
    q_{n+1}=\textbf{A} q_{n}. 
\end{equation}
However, $\textbf{Q}$ and $\textbf{A}$ are the full snapshot size ($M \times N$ and $M \times M$, respectively).
For computational economy, the snapshots and dynamic operator are projected onto POD modes, obtained by first splitting and shifting the snapshot matrix into $ \textbf{Q}_1= [q_1,...,q_{N-1}]$ and $\textbf{Q}_2=[q_2,...,q_{N}]$,
and then performing the singular value decomposition
 \begin{equation}\label{eqn:SVD}
 \textbf{Q}_1=\textbf{U}\boldsymbol{\Sigma} \textbf{V}^{*},
 \end{equation}
 to produce POD modes $\textbf{U}$, which are used to calculate the reduced DMD operator $\tilde{\textbf{A}}$,
 \begin{equation}\label{eqn:Atilde}
 \tilde{\textbf{A}}=\textbf{U}^{*}\textbf{A}\textbf{U}=\textbf{U}^{*}\textbf{Q}_2\textbf{V}\boldsymbol{\Sigma^{-1}}.
 \end{equation}
The $\tilde{\textbf{A}}$ matrix is then of size $N\times N$.
Note that a POD mode truncation can be used to further reduce or stabilize $\tilde{\textbf{A}}$.
This becomes necessary for the filtered schlieren data as discussed in Appendix~\ref{secn:append}.

Reduced snapshots, $\tilde{\textbf{Q}}=[\tilde{q}_1,...,\tilde{q}_N]$, are projected onto POD modes
\begin{equation}
    \tilde{\textbf{Q}}=\textbf{U}^{-1} \textbf{Q},
\end{equation}
and consistently evolved forward in time, reproducing the input solution.
The stability and receptivity analysis is enabled by the addition of a forcing term:
\begin{equation}\label{eqn:dmd_Forcingl}
    \tilde{q}^{'}_{n+1}=\tilde{\textbf{A}} \tilde{q}_{n} + \tilde{f}_{n}.
\end{equation}
As shown in Fig.~\ref{fig:DMD_method}(b), user-defined forcing snapshots, $\textbf{F}=[f_1,...,f_N]$ provide the forcing term and are of consistent size with the schlieren snapshots because they are  projected onto the reduced subspace using the POD modes of the flow data,
\begin{equation}
    \tilde{\textbf{F}}=[\tilde{f}_1,...,\tilde{f}_N]=\textbf{U}^{-1} \textbf{F}.
\end{equation}

After the DMD operator, schlieren, and forcing snapshots have been reduced, the system is iterated forward in time under the influence of forcing, according to Eqn.~\ref{eqn:dmd_Forcingl} and shown in Fig.~\ref{fig:DMD_method}(c).
The response snapshots, $\tilde{\textbf{Q}}'=[\tilde{q}^{'}_1,...,\tilde{q}^{'}_N]$, are collected and expanded back to the full sized domain for observation by
\begin{equation}\label{eqn:expansion}
    \textbf{Q}'=\textbf{U}\tilde{\textbf{Q}}'.
\end{equation}
Apart from directly observing the forced response snapshots, the energy amplification is measured using gain metrics that compare the L2 energy norm of the response to the forcing.
For parametric studies, the gain is more efficiently calculated in the reduced space to avoid the expansion step (Eqn.~\ref{eqn:expansion}).
The total gain over all snapshots is defined as,
\begin{equation}\label{eqn:gain}
\sigma=\frac{||\tilde{\textbf{Q}}'||_2}{||\tilde{\textbf{F}}||_2},
\end{equation}
and is used for the harmonic forcing analysis.
A transient gain is also defined for individual snapshots
\begin{equation}
\sigma(t)=\frac{||{\tilde{q}(t)}'||_2}{||\tilde{\textbf{F}}||_2}.
\end{equation}
For impulse forcing, this gain is defined with respect to the initial forcing snapshot $f_o$.
The behavior of decaying amplitude perturbations may be exaggerated by using the transient gain to normalize the impulse response $q(t)'/\sigma(t)$ as discussed in Section~\ref{secn:DMD_ROM}.


In previous studies using the DMD-ROM \citep{Stahl_2024_DMD}, the forcing was applied to the unsteady flow, emulating a simpler analog of the synchronized LES approach of \citet{adler2018dynamic}.
From an alternative perspective of a perturbed basic or equilibrium state, the forcing would be applied to the mean-flow to better isolate the transient response, similar to mean-flow perturbation (MFP) analysis \citep{RAJ18}.
This latter perspective is reproduced here with the DMD-ROM by replacing the initial condition snapshot with the mean-flow.
However, because DMD is able to distinguish the mean-flow as a mode without a growth rate or frequency, the mean-flow implicitly does not evolve with time. 
As such, the dynamic operator only evolves the perturbations in time without contributions of the mean flow, which would be subtracted anyway in the case of MFP. 
Due to its redundancy, here the mean-flow initial condition is replaced by a zero-vector snapshot base flow. 

The forcing snapshots are populated with impulse or harmonic forcing at locations just upstream of the splitter-plate trailing edge.
The impulse is applied only at the first time-step, while the harmonic forcing lasts the entire duration of the schlieren video, $f(t)=A\cos(\omega t)$.
The impulse forcing thus targets the growth over a shorter time horizon, while the harmonic forcing investigates the asymptotic, statistically-stationary response.
Note that in contrast to typical stability methods that necessarily consider small perturbations, the DMD-ROM forcing amplitude tests conducted by \cite{Stahl_2024_DMD} revealed that larger amplitudes were more effective in eliciting the relative linear gain rates across forcing parameters (i.e. receptivity).
This is primarily dependent on the size of the perturbations relative to the fluctuations of the unsteady flow.
Here, the perturbation response is calculated without the comparatively larger unsteady fluctuations or mean flow.
Under these conditions, the output response is simply scaled by the input amplitude due to the linearity of the DMD-ROM; this scaling is accounted for by the normalization of gain in Eqn.~\ref{eqn:gain}.
Therefore, the results do not change as a function of forcing amplitude, which is arbitrarily set to unity for all cases.






\subsection{Experiments and schlieren data} \label{subsec:schlieren}

\begin{figure}
   \centerline{\includegraphics[width=01\textwidth]{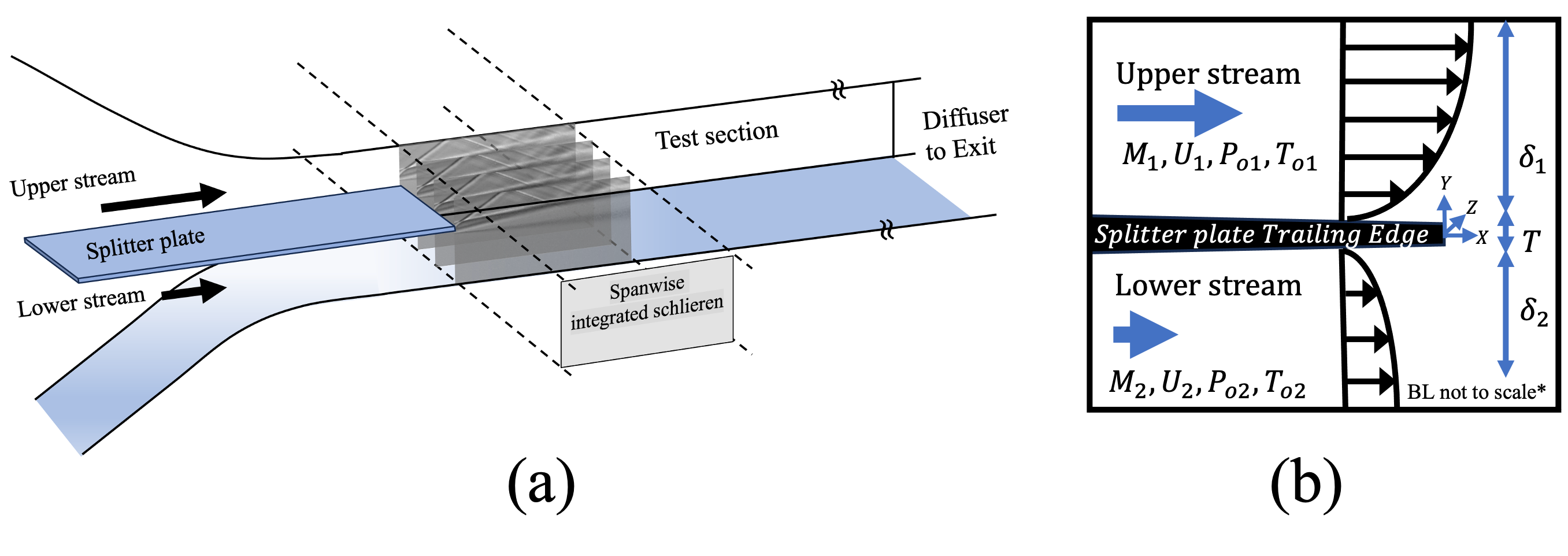}}
   \caption{(a) Wind tunnel configuration and test section. (b) Splitter plate, incoming boundary-layers, and flow conditions. Not to scale. 
   }
\label{fig:flow_intro}
\end{figure}

Turbulent shear-layer measurements are used from \citet{KEVIN20B}, who obtained results at several test conditions reflecting different levels of compressibility effects; these are documented at
https://wiki.illinois.edu/wiki/display/NCSLF.
Case~5, corresponding to $M_c=0.88$ ($M_1=2.461$ and $M_2=0.175$) is chosen for reasons discussed in Section~\ref{secn:intro_SL}.
A schematic of the test apparatus is introduced in Fig.~\ref{fig:flow_intro}(a) with the splitter plate, incoming boundary layers and shear-layer configuration isolated in (b).
For reference, the flow conditions are reproduced in Table~\ref{tab:flowcond} with boundary layer characteristics listed in Table~\ref{tab:BL}.
The splitter plate thickness relative to the boundary-layer height is an influential parameter for mixing layer dynamics, since it controls the relative importance of shedding versus Kelvin-Helmholtz (K-H) instabilities; a recent effort by \citet{Stack2019} systematically highlights the principal effects as the plate thickness is changed.
Here, the splitter plate tapers to a thin trailing edge with a thickness of $H=0.5~mm$, and marks the coordinate system origin for the streamwise ($X$), wall-normal ($Y$), and spanwise ($Z$) directions. 
The ratio of the splitter-plate thickness to boundary-layer height is approximately $H/\delta_1=1/6$ and $H/\delta_2=1/8$ for the upper and lower streams respectively; this suggests immediate mixing of small-scale structures, as evidenced in Fig.~\ref{fig:schlieren_intro}.
The spanwise extent of the configuration is $127~mm$, which, compared to the thin side-wall boundary layers, establishes spanwise homogeneous flow in the center of the duct where schlieren line-of-sight integration is focused.

\begin{table}
\centering
\caption{Flow conditions of upper and lower streams}
\label{tab:flowcond}
\begin{tabular}{llllll}
\hline
Stream    & Mach  & $P_0$ (kPa) & $T_0$ (K) & $U$ (m/s) & Re ($m^{-1}$) \\
\hline
Upper (1) & 2.461 & 778.10     & 289.25   & 564.16  & 8.455E+07                  \\
Lower (2) & 0.175 & 50.95      & 292.58   & 59.76   & 1.983E+06                 
\end{tabular}
\end{table}

\begin{table}
\centering
\caption{Boundary layer conditions on splitter plate}
\label{tab:BL}
\begin{tabular}{lllllll}
\hline
Stream    & $\delta$ (mm)  & $\delta^*$ (mm) & $\theta$ (mm) & $H$ & $C_f$ & $\Pi$ \\
\hline
Upper (1) & 3.108 & 0.582     & 0.419   & 1.390  & 0.000871   & 1.6398               \\
Lower (2) & 4.007 & 0.510     & 0.390   & 1.308   & 0.004618   & 0.2008              
\end{tabular}
\end{table}


Optical measurements were obtained across the entire $127 \times 127 \times 762~mm$ test section, as depicted in Fig.~\ref{fig:flow_intro}(a).
Due to spatio-temporal resolution considerations, the test section is partitioned into multiple smaller subsections.
\begin{figure}
   \centerline{\includegraphics[width=01\textwidth]{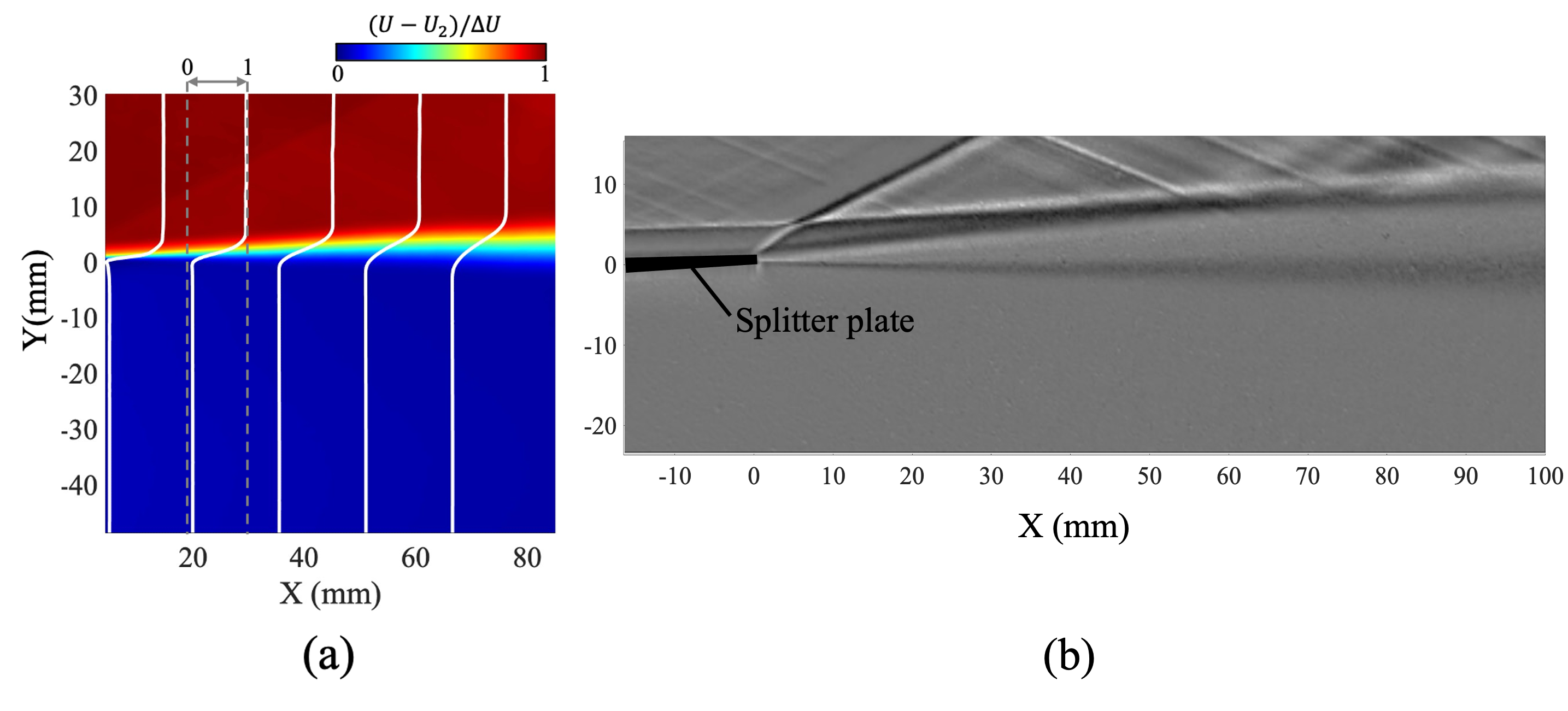}}

  \caption{(a) Time-mean velocity field  from PIV data. (b) Time mean schlieren image.}
\label{fig:mean}
\end{figure}
For reference, PIV measurements of the time-mean velocity field in the subdomain immediately downstream of the splitter plate is presented in Fig.~\ref{fig:mean}(a), showing the mixing layer region and velocity profiles at multiple streamwise locations.
Details of the PIV setup with in-depth analysis of the velocity field and Reynolds stresses may be found in \citet{KEVIN20B}.
More pertinent here are schlieren measurements, which use the angular deflection of collated light passing through the spanwise direction of the flow to record pixel intensities modulated by streamwise density gradients. 
The 
time-mean schlieren image is shown in Fig.~\ref{fig:mean}(b), better illustrating the shock structures in the upper stream. 
In this work, the subdomain closest to the splitter plate is used, which has slightly different dimensions for PIV and schlieren windows. 
The schlieren field of view extends \textbf{$126.5 \times 39.5$} mm around the splitter plate and is resolved with $384 \times 128$ pixels.
The smaller region  affords a higher sampling rate of $F_s=124~kHz$, which is sufficient to calculate the time-derivatives for the irrotational filtering and DMD-ROM analysis. 
$2{,}500$ images, spanning $20.2~ms$, are used to form the models.
Since the interest is in the transient response of instabilities, time is reported in non-dimensional form throughout the paper by $t^*=tF_s$.

\section{Spectral analysis}\label{secn:SPOD}
\begin{figure}
\centering
\includegraphics[width=1\textwidth]{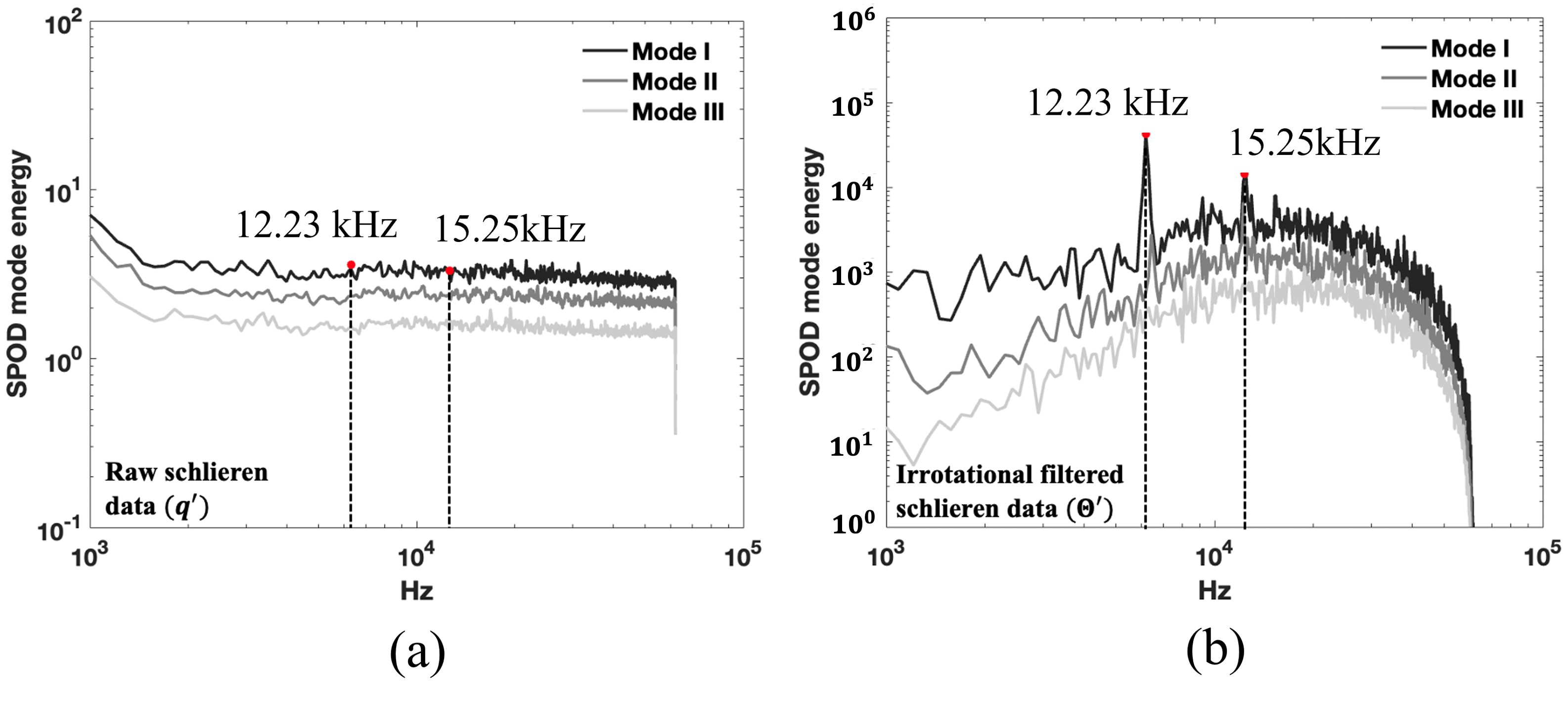}
\caption{SPOD spectra for the (a) raw and (b) filtered schlieren data.}
\label{fig:SPOD_SPECTRA}
\end{figure}

Prior to investigating the transient forcing and response, a conventional spectral analysis is conducted to demonstrate the advantages of the filtered schlieren data over the raw values. 
SPOD is used with the block Welch's formulation of \citet{Schmidt2020}.
The $2{,}500$ images are partitioned into three blocks with 50\% overlap, resolving a frequency range of $150$ to $62{,}000$ Hz.
The resulting spectra of the first three modes are shown in Fig.~\ref{fig:SPOD_SPECTRA} for the (a) raw and (b) filtered data.
Note the energy scales compared across datasets is arbitrary, since the relative trends across frequencies are the more crucial quantities of interest.
The (a) raw schlieren is dominated by ultra low frequencies; this bias is likely
not physical and is attributed to camera effects.
In the dynamically significant range beyond $10^3$ Hz, the constant broadband spectra yield little insight into which modes are important. 
In contrast, the (b) filtered schlieren shows the irrotational energy is biased towards a higher frequency band, and is clearly dominated by several tones.
The largest tones ($12.23$ and $15.25~kHz)$ are selected for further analysis.

\begin{figure}
\centering
\includegraphics[width=1\textwidth]{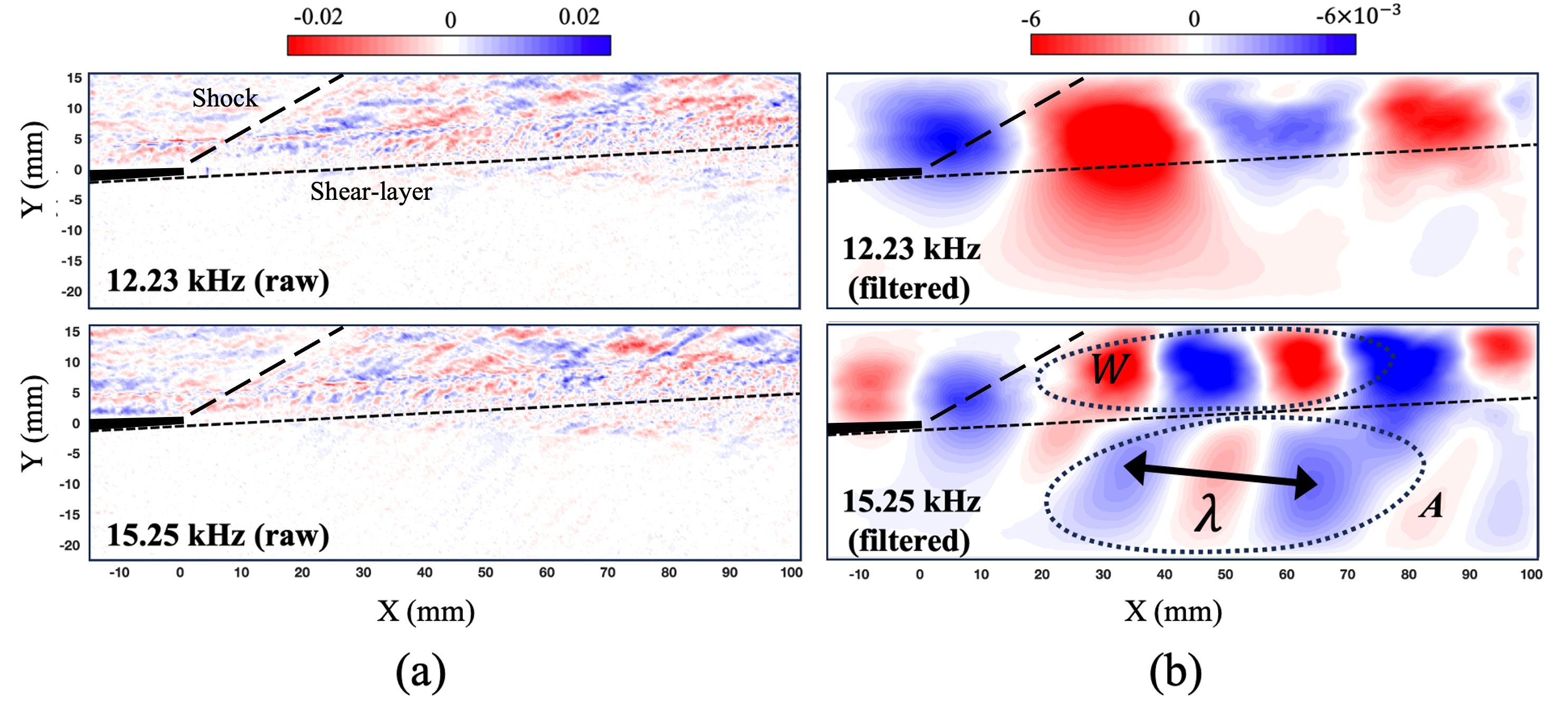}
\caption{SPOD modes for the (a) raw and (b) filtered schlieren data at the peak acoustic tones (12.23 and 15.25 kHz). The dashed lines represent the approximate locations of the shock and the shear-layer extended from the splitter plate.}
\label{fig:SPOD_modes}
\end{figure}

The corresponding SPOD modes of the identified tones are presented in Fig.~\ref{fig:SPOD_modes}, again comparing the (a) raw and (b) filtered schlieren.
The black dotted lines reference the shock and shear-layer line extending between supersonic and subsonic streams, as obtained from Fig.~\ref{fig:mean}(b), of which the supersonic stream contains stronger fluctuations. 
Here, the differences between the data sets are clear; the (a) raw schlieren modes display low coherence associated with the mixing of smaller turbulent structures, primarily in the upper stream, while the (b) filtered schlieren modes feature larger and well-defined wavepacket structures.
For the most energetic $12.23~kHz$ mode, the largest irrotational structures occur in the supersonic stream.
The higher frequency $15.25~kHz$ mode displays smaller waves, with a distinct change in wavepacket structure before and after the shock.
Also of significance are the acoustic waves propagating in the lower stream that are not evident in the raw modes.
The filtered schlieren SPOD modes are very revealing in terms of isolating the general structure of upper-stream wavepackets ($W$) and lower stream acoustics ($A$); and of note for later, the acoustic wavelength $\lambda$.
However, the convective phenomenon that relates these different wavepacket structures across all frequencies is not entirely clear in SPOD modes, motivating the transient impulse analysis to further distinguish the growth and evolution of acoustic waves. 


\section{Forced response analysis}\label{secn:DMD_ROM}


The forced response analyses are first conducted with an impulse applied across both boundary-layers simultaneously to examine the overall stability properties of the shear-layer.
This is followed by two more precise receptivity studies, where specific locations are perturbed independently with impulse or harmonic forcing.
For the DMD-ROM method, both forcing types (see Section~\ref{secn:DMD-ROM}) are implemented by constructing empty snapshots and populating individual pixels in the wall-normal direction of the splitter-plate with the prescribed forcing function over the duration of the video.

\subsection{Impulse forcing across both boundary layers} \label{sec:impulse}
\begin{figure}
\centering
\includegraphics[width=1\textwidth]{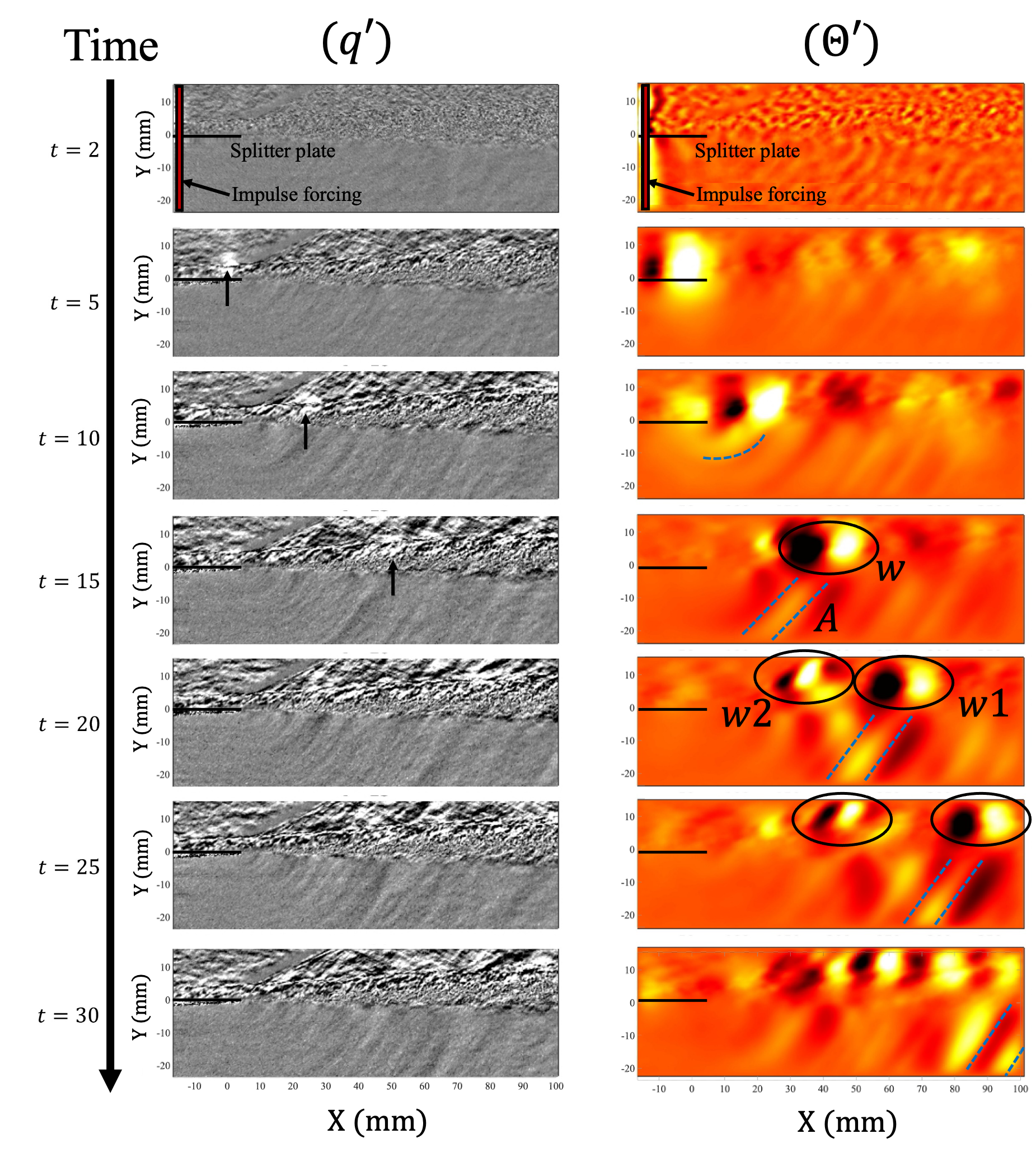}
\caption{Impulse response comparison of the raw (left) and filtered (right) schlieren, progressing in time from top to bottom. The impulse location in shown in the first frame. Wavepacket ($W$) and acoustic ($A$) structures are identified convecting downstream.}
\label{fig:impulse_compare}
\end{figure}

Figure~\ref{fig:impulse_compare} displays the impulse response for the situation where
the disturbance is applied across a single vertical column of pixels at $X=-12~mm$ (outlined at time $t^*=2$) such that it perturbs both boundary-layers simultaneously.
Results from the raw and irrotational filtered schlieren data are presented in the left and right columns, respectively, and scaled by the transient gain, as discussed in Section~\ref{secn:DMD-ROM}, to better visualize the structures in each frame.
The signature of the initial disturbance is observed around this location and induces a relatively quiescent background of small-scale noise; an artifact of the DMD model which is quickly damped out.
The disturbance progresses in time from top to bottom, growing into a large wavepacket structure (\textit{w}) that is noticeably more coherent in the filtered data.
By time $t^*=20$, it is clear that the wavepacket in the upper stream disperses when crossing the shock (\textit{w1} and \textit{w2}) and two different speeds become apparent.
In contrast, the initial impulse disturbance in the lower stream is much weaker compared to the larger waves in the upper stream.
However, an acoustic wave ($A$) generated later at the trailing edge of the splitter plate ($t^*=10$) propagates through this subsonic region, and is followed by a series of weaker waves emitted from the shock interaction.
Such acoustic waves that propagate through the subsonic region are evidently related to the corresponding supersonic convection of hydrodynamic structures in the upper stream \citep{Hall1993}.
Nonetheless, as the raw schlieren response progresses in time, these disturbed structures are barely discernible as they become indistinguishable in the turbulent mixing layer.

\begin{figure}
\centering
\includegraphics[width=1\textwidth]{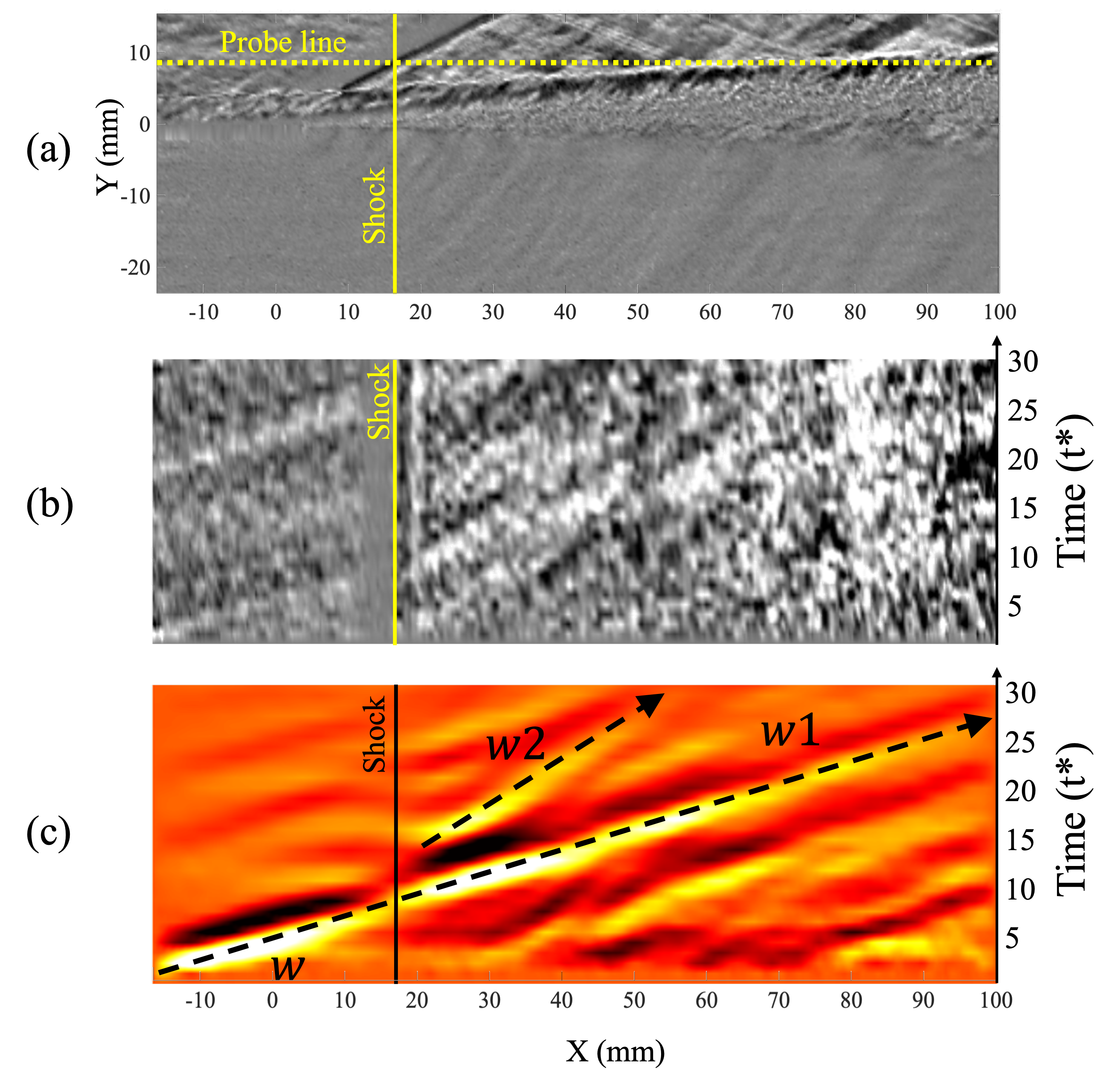}
\caption{(a) Data extraction location (dashed line) in the upper stream. Corresponding space-time plots from the (b) raw and (c) filtered schlieren impulse responses. The initial wavepacket $w$, bifurcates across the shock, maintaining speed and structure with $w1$ and producing a slower wavepacket $w2$. }
\label{fig:spacetime_comapare}
\end{figure}

The impulse wavepackets are further investigated with space-time plots of the response in Fig.~\ref{fig:spacetime_comapare}.
The data is extracted from pixels located on the dashed line in Fig.~\ref{fig:spacetime_comapare}(a) to capture the wavepacket-shock interaction region.
Space-time plots are compared for the (b) raw and (c) filtered schlieren results, again demonstrating a clearer representation of the wave propagation pattern in the irrotational data.
As time progresses, downstream propagating waves are interpreted as streaks traveling up and to the right, with a speed corresponding to the slope of the line.
Examining (c), the incipient wavepacket ($w$) has a rightward running path that maintains a constant slope after the shock ($w1$), corresponding to an approximate speed of $535 m/s$.
The second wavepacket ($w2$) is generated after the shock and has a steeper slope, which corresponds to a slower $306 m/s$.

\begin{figure}
\centering
\includegraphics[width=1\textwidth]{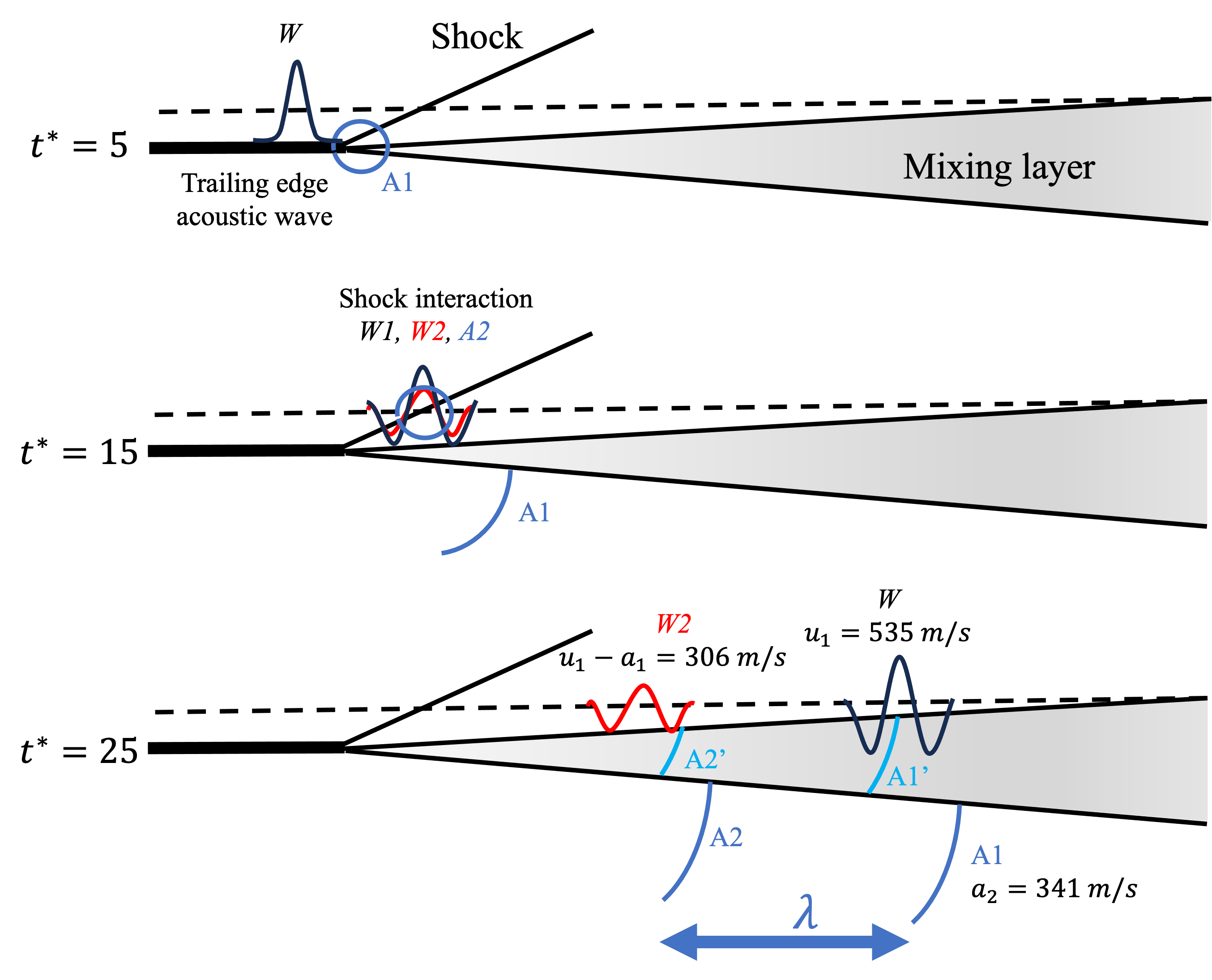}
\caption{Schematic of the shear-layer instability and dispersion of wavepackets across the shock. The initial wavepacket $w1$ emits acoustic wave $A1$ at the splitter plate trailing edge. The wavepacket-shock interaction produces the slow acoustic wavepacket $w2$ and acoustic emission $A2$.}
\label{fig:imp2}
\end{figure}

Figure~\ref{fig:imp2} presents a generalized illustration of the acoustic wavepacket dynamics produced by the shear-layer instability.
The dashed line represents the location where space-time plots were extracted to compute the wavepacket velocities shown in the diagram.
Starting at $t^*=5$, the incoming disturbance along the upper boundary-layer reaches the trailing edge of the splitter plate, emitting the acoustic wave $A_1$ in the lower stream.
Further downstream, at approximately $t^*=15$, the disturbance interacts with the shock, bifurcating the upstream wavepacket $w1$ and $w2$.

It is well established that such a shock interaction results in a complicated dispersion of vortical, acoustic, and entropic waves, with varying phase speeds and orientations based on the shock angle and strength \citep{duck1995interaction}.
The filtered schlieren data isolates the acoustic components of this process.
Given the measured velocities and nominal flow conditions in the upper stream ($M_1=2.46, U_1=564 m/s$ and $a_1=229 m/s$, before the shock), wavepacket $w1$ is limited by the supersonic flow speed $u_1=535 m/s$, while $w2$ matches the slow acoustic wave speed $u_1-a_1=306 m/s$.
This interaction also generates an acoustic wavefront $A2$, distinguished from $w1$ for its propagation out of the supersonic stream and into the subsonic flow. 
As shown at $t^*=25$, dispersion effects from the variable Mach number across the shear-layer causes refracted portions of the acoustic wave ($A_1'$ and $A_2'$) to travel at slower speeds when passing through supersonic flow, although, this is more difficult to measure with limited pixel resolution in the mixing region.
Finally, the time delay between the production of the trailing edge acoustic wave $A1$ and the slower shock acoustic wave $A2$, defines the distance $\lambda$, observed earlier, for instance in the SPOD mode wavelength of Fig.~\ref{fig:SPOD_modes}(b).
 \subsection{Impulse location study} \label{sec:impulselocation}
\begin{figure}
\centering
\includegraphics[width=1\textwidth]{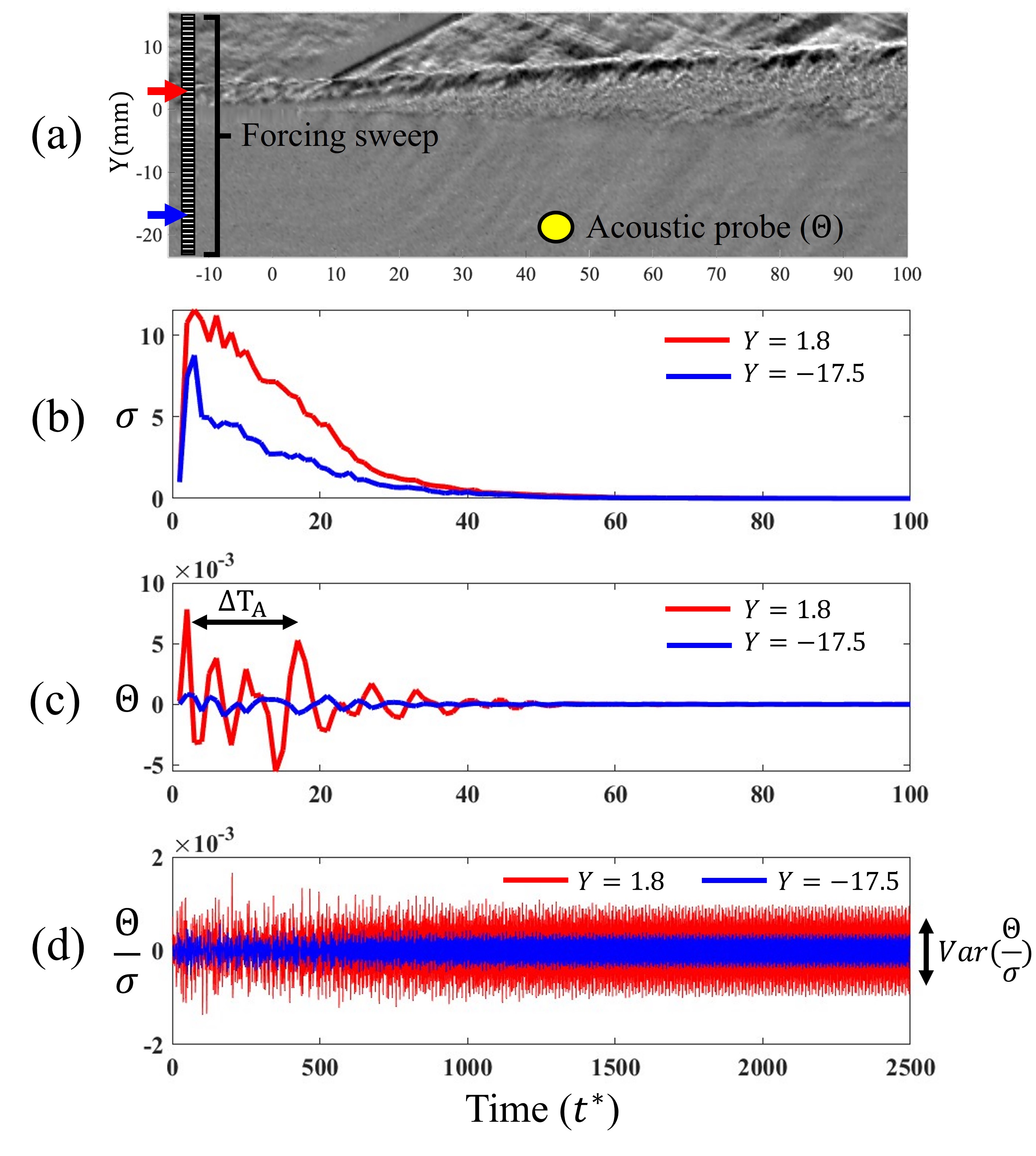}
\caption{(a) Parametric impulse forcing locations and acoustic probe location where the response is measured. All analyses are conducted on the irrotational filtered variable $\Theta$. The most ($Y=1.8$) and least ($Y=-17.5$) amplified responses are compared. (b) Transient gain, $\sigma$, measures the energy growth and decay across the entire domain. (c) Acoustic probe fluctuations $\Theta$ are used to measure the time lag $\Delta T$ between successive acoustic waves. (d) The acoustic signal is normalized by the transient gain, $\Theta/\sigma$ to better compare the asymptotic energy at the probe location, as measured by the variance of the fluctuations.  }
\label{fig:impulse_probe}
\end{figure}

The time-delay between waves in the acoustic field may be explicitly measured by using isolated disturbances to further analyze the response, including the value of $\lambda$.
Individual pixels are perturbed along the same vertical column as before to find the most sensitive locations in each stream. 
The acoustic response is measured by a probe in the lower stream at $X=45~mm$, identified by the yellow circle in Fig.~\ref{fig:impulse_probe}(a).
The instantaneous raw schlieren image in Fig.~\ref{fig:impulse_probe}(a) is only shown for reference, while the analysis is conducted on the irrotational flow, $\Theta'$, to measure the acoustic output produced by each disturbance location. 
To illustrate the disparate effects of forcing location, the most and least amplified responses from the upper (red, $Y=1.8$) and lower (blue, $Y=-17.5$) streams are compared.  

The impulse response is measured in three ways in Fig.~\ref{fig:impulse_probe}: the (b) transient gain encompassing the energy of the entire flow field, the (c) acoustic probe signal, and the (d) probe signal normalized by the transient gain.
Examining the transient gain of (b), the initial energy amplification of the convective instability is followed by asymptotic decay.
Both forcing locations experience their peak growth at $t^*=3$ but have different decay rates.
The probe signal in~(c) illustrates the initial response of acoustic fluctuations, representing the passage of waves, which similarly decay over time.
For the $Y=1.8$ input forcing, the time between the two largest peaks is $\Delta T_A=15=0.121 ms$ and corresponds to the lag between acoustic waves $A1$ and $A2$.
Given the speed of sound measured in the experiments, this yields an instantaneous acoustic wave distance $\lambda=a_2 \Delta T_A = 41~mm$.
This is comparable to the characteristic wavelength observed in the $15.25~kHz$ SPOD mode of Fig.~\ref{fig:SPOD_modes}(b) in this probe region.
The acoustic signal in (d), normalized by the transient gain, provides a clearer comparison of fluctuation size over time, resulting in a periodic signal as the transients decay.
The asymptotic frequency is $31~kHz$, which is a harmonic of the peak SPOD tone; this frequency represents an artifact of the slowest decaying mode in the DMD model. 
Regardless, the transient normalization is useful because, for the same forcing amplitude, the relative energy output is preserved over time, as measured by the variance of the normalized signal (d), which is utilized in the upcoming parametric study.  

\begin{figure}
\centering
\includegraphics[width=1\textwidth]{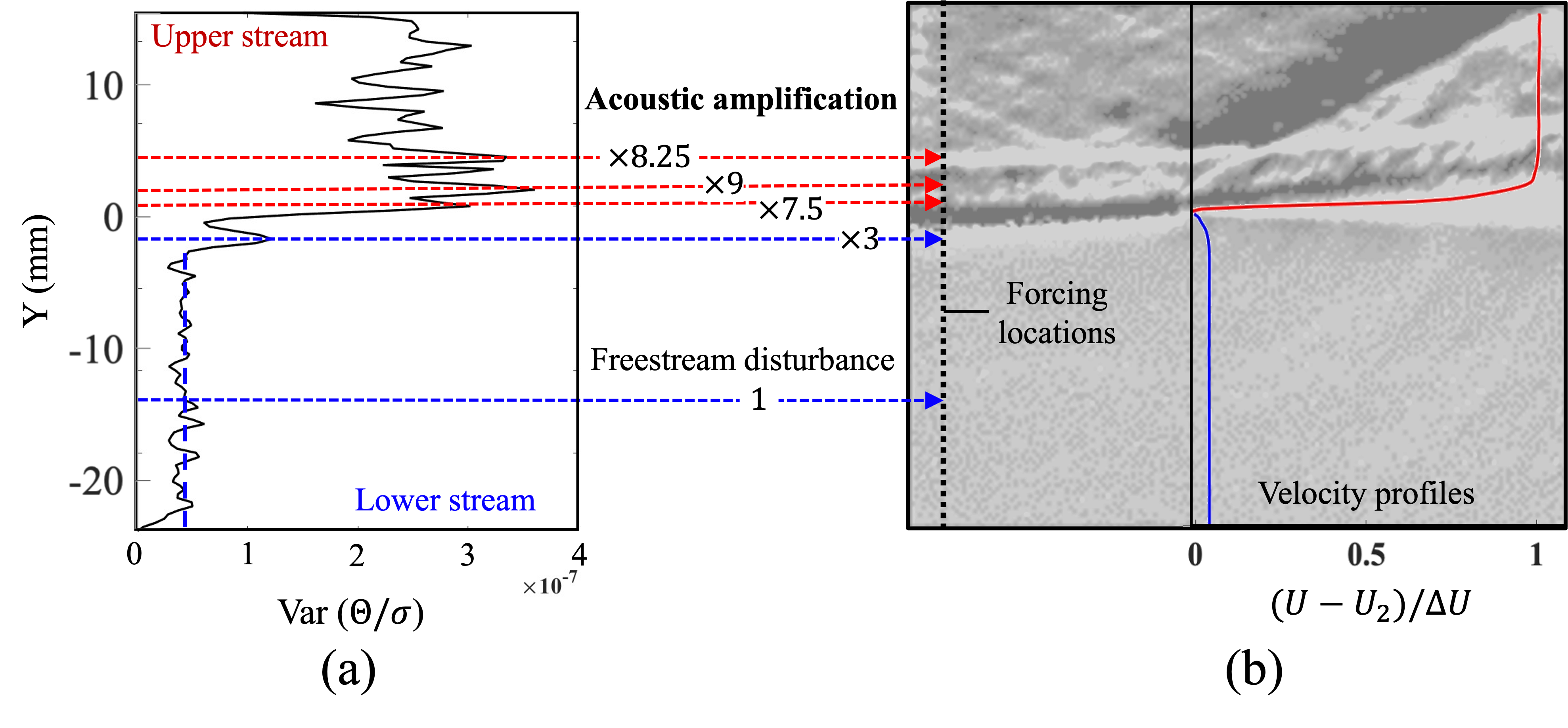}
\caption{(a) Variance of the normalized acoustic signals (measured at the probe in Fig.~\ref{fig:impulse_probe}(a)) as a function of wall-normal impulse location. The (b) schlieren image and streamwise velocity profile provide reference for these locations. The variance peaks are identified and normalized relative to the average of the lower stream, showing significant multiples of acoustic output in the upper stream.  }
\label{fig:impulse_variance}
\end{figure}

The amplitude response of all individual forcing locations is compared using the probe signal variance in Fig.~\ref{fig:impulse_variance}(a).
This metric effectively determines the relative acoustic energy output of disturbances placed along the upper (red) and lower (blue) streams.
These input locations are mapped to the flow field in Fig.~\ref{fig:impulse_variance}(b), depicting a schlieren image overlaid with time-mean streamwise velocity profiles, taken from PIV data immediately after the splitter plate.
The overall trend in variance shows several peaks in the upper stream that are much larger than the lower stream, which is nearly constant apart from the boundary layer region.
The average lower stream variance (vertical dashed line in (a))
of acoustic disturbances in the subsonic free stream is not produced from the splitter plate and is more representative of ambient wind-tunnel noise.

This background noise is used to normalize the relative energy amplification of other peak locations, marked with dashed lines in Fig.~\ref{fig:impulse_variance}. 
These peaks correspond to the lower-stream boundary layer ($Y=-1.8~mm$), and upper-stream boundary layer ($Y=0.6,1.8,$ and $4.0$ mm).
The obvious conclusion is that the maximum acoustic energy is generated from perturbations in the upper-stream, outer boundary-layer region and is about nine times larger than subsonic free stream disturbances.
By comparison, the maximum energy in the lower stream boundary-layer is only three times greater than free stream disturbances.
These results illustrate how the combined post-processing techniques can yield relative amplification data from qualitative schlieren data, and are consistent with receptivity studies showing higher-speed flow is more receptive at the boundary-layer edge to freestream disturbances which can sustain growth downstream \citep{schmid2001stabilitybook}.




\subsection{Harmonic forcing sweep} \label{sec:harmonicforcingsweep}
The receptivity and stability properties of the shear-layer are now investigated with harmonic forcing. 
This parametric study tests a range of forcing frequencies ($0$ to $60$ kHz) at the same pixel locations as the previous analysis.
The same DMD operator, forcing amplitude, and initial conditions are used as the impulse study.

\begin{figure}
\centering
\includegraphics[width=1\textwidth]{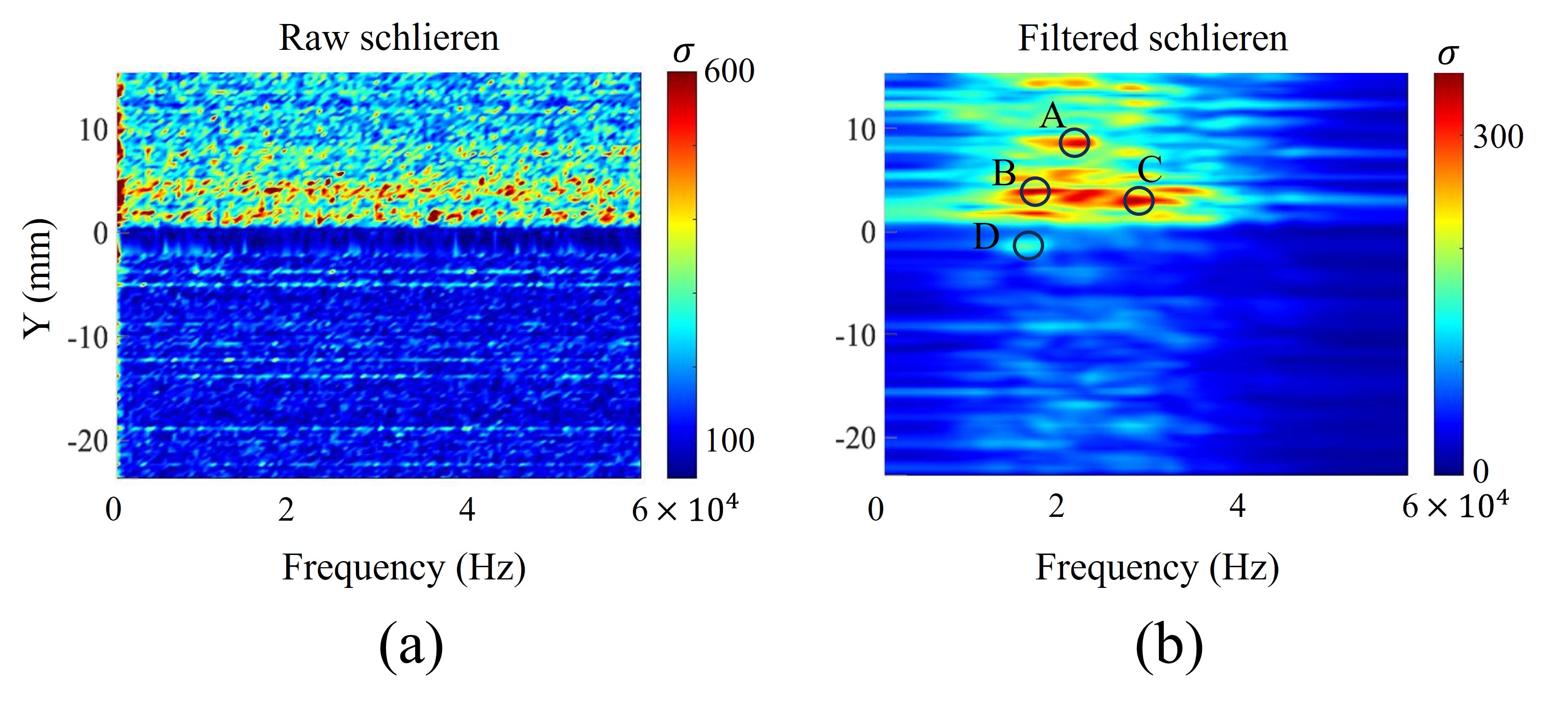}
\caption{Harmonic forcing gain $\sigma$ of the (a) raw and (b) acoustic filtered schlieren data as a function of forcing pixel location in the splitter-plate boundary layer.  }
\label{fig:freq_sweep}
\end{figure}

The total gain across all response snapshots is first presented in Fig~\ref{fig:freq_sweep}, comparing (a) raw and (b) irrotational schlieren data.
Both gain plots affirm that the supersonic stream is more receptive to amplifying disturbances than the lower stream, for nearly all frequencies.
The raw schlieren, (a), is generally noisy with flat broadband spectra.
As with the SPOD results, this observation does not provide much insight into the most important frequencies. 
In contrast, the filtered schlieren analysis, (b), shows several prominent tones in the supersonic stream in the $20-40~kHz$ range; a few of these are selected for further analysis and labeled A-D.
The most receptive conditions (B) occur for a frequency of $17.5~kHz$ at $Y=3.3$ near the outer edge of the upper stream boundary-layer, where the velocity gradients change rapidly.
The next most amplified conditions, (A) and (C) respectively, are also at nearby frequencies and located in the upper stream. 
Although the lower stream gain is much weaker for (D), a distinct frequency of $16~kHz$ is observed and also selected.

\begin{figure}
\centering
\includegraphics[width=1\textwidth]{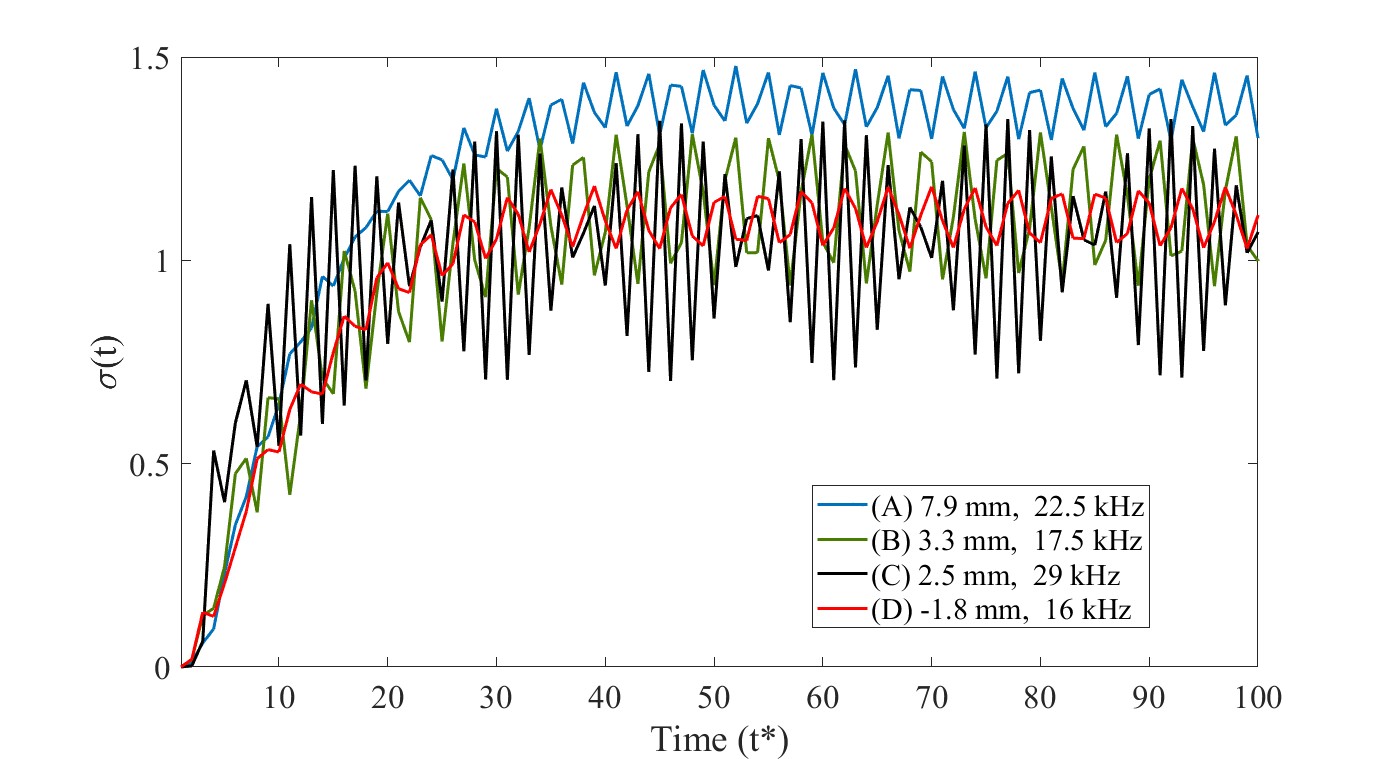}
\caption{Transient gain of the harmonic forcing response for the four cases identified in Fig.~\ref{fig:freq_sweep}.}
\label{fig:freq_transient}
\end{figure}

The transient gains at the four identified frequencies and locations are presented in Fig.~\ref{fig:freq_transient} for the acoustically filtered results.
All forcing inputs show initial transients before asymptoting to the steady state response. 
The growth period lasts much longer than the previous impulse results since it is sustained by the continual input of energy. 
Note that some cases have a higher transient gain $\sigma(t)$ than those  identified with larger total gains $\sigma$, which is a measure of the fluctuation size.
For example, (A) reaches an average, asymptotic gain of 1.4, but the fluctuations of (B) are larger, despite it having a lower asymptotic average gain of 1.1.
The asymptotic gain oscillates at frequencies that are much higher than the input forcing. 
The most surprising result is the resonant beating envelope for case (C), suggesting a superposition of multiple spatial and temporal waves in the flow that interact with the forcing frequency.

\begin{figure}
\centering
\includegraphics[width=1\textwidth]{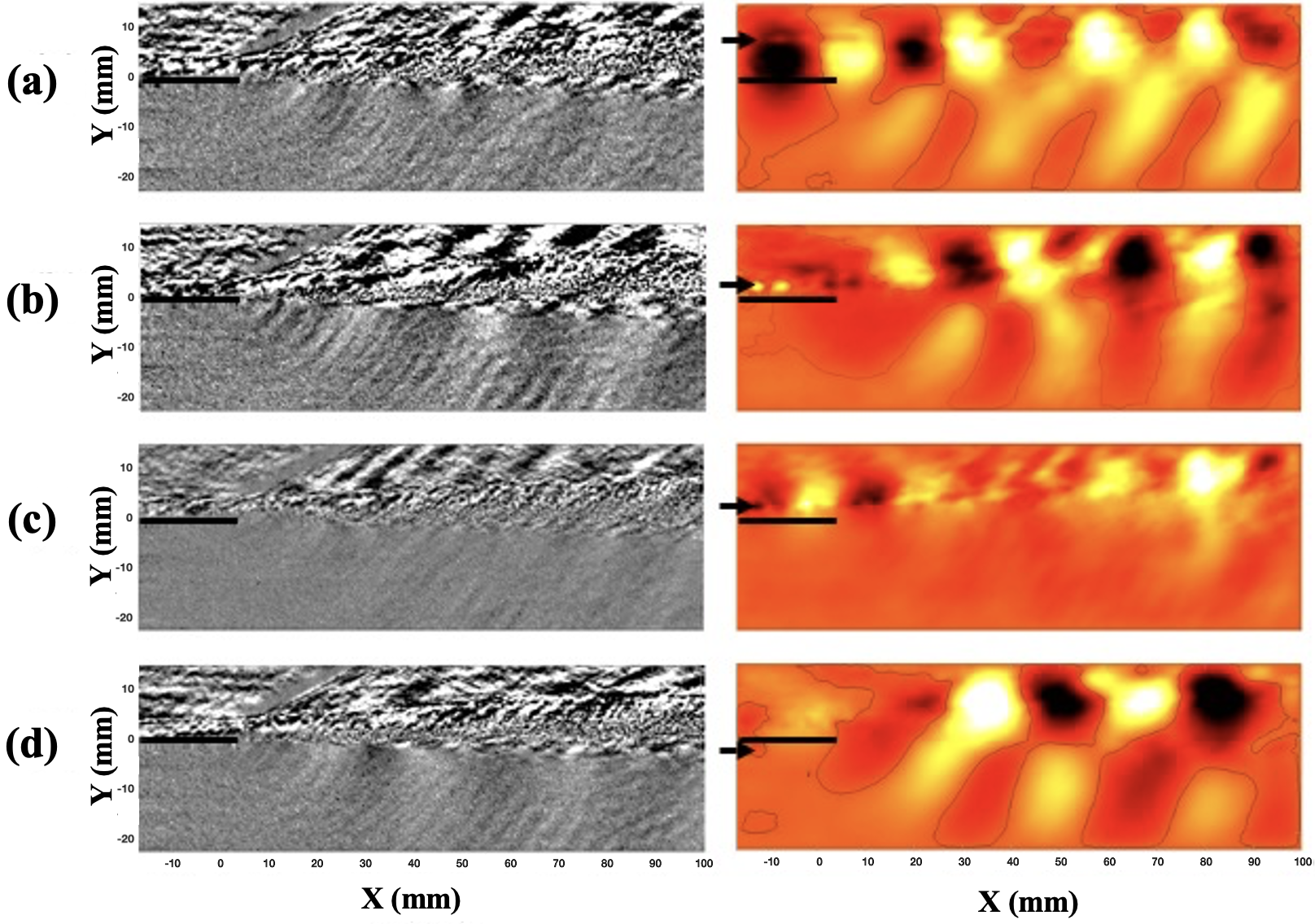}
\caption{Instantaneous snapshots of the harmonic forcing response showing the raw (left) and filtered (right) schlieren results for the forcing conditions at (a) $Y=9.9~mm, 22.5~kHz$, (b) $Y=3.3~mm, 17.5~kHz$, (c) $Y=2.5~mm, 29~kHz$, and (d) $Y=-1.8~mm, 16~kHz$. }
\label{fig:freq_respons_snaps}
\end{figure}

More physically insightful than the transient gain is the actual flow response, as captured by instantaneous snapshots during the asymptotic stage of the forcing.
The instances selected represent the salient dynamics throughout the periodic cycle corresponding to the forcing frequency. 
Figure~\ref{fig:freq_respons_snaps} shows both the raw schlieren (left column) and filtered schlieren (right column) under forcing parameters (A) through (D).
The raw schlieren generally shows small-scale turbulent structures in the shear-layer, while the filtered images reveal larger wavepacket structures and acoustic radiation that emphasize longer wavelengths.
Also noticeable in the filtered results are the incipient disturbances near the forcing locations, marked by arrows.
In the case of~(A) and (C), these are accompanied by larger wavepackets above the splitter plate, while~(B) and~(D) display structures primarily downstream of the shock.  
The acoustic radiation wavelength for Case~(A) is consistent in both the upper and lower stream.
Case~(B), which is the most energetic response, has a more complicated structure, particularly around the shock region.
Case~(C) demonstrates that this forcing is receptive, but not prone, to generating acoustic radiation, which is lacking in the lower stream.
This suggests potential forcing conditions that the flow could ``lock-on'' to  and without amplifying emitted noise.
Finally, case (D) shows the forcing in the lower stream has a unique angle of acoustic propagation compared to the other cases. 
This case also demonstrates that under harmonic forcing conditions, the DMD-ROM response is closely related to spectral modes, as compared by  similar SPOD modes of Fig~\ref{fig:SPOD_modes}(b) at the nearby $15.25~kHz$ frequency.
In this case however, the DMD-ROM can target the precise locations that are sensitive to these frequencies, suggesting the integral role of the lower stream in noise generation.


\section{Conclusions}\label{secn:conclusion}
A novel combination of two post-processing methods to analyze time-resolved schlieren data was advanced and applied to obtain acoustic receptivity and stability insights.
A transonic shear-layer characterized by $M=2.46$ and $M=0.175$ upper and lower streams, respectively, and convective Mach number $M_c=0.88$ is used as a testbed.
The first step is a schlieren-specific analog to momentum potential theory, which filters the irrotational component of the flow, thus isolating acoustic wavepackets and propagation within the turbulence.
Spectral proper orthogonal decomposition demonstrates the advantages of the filtering; whereas the raw schlieren shows only broadband spectra, the filtered results very clearly display dominant tones.
The second component is a dynamic mode decomposition reduced order model (DMD-ROM) used to test the forced response of the shear-layer.
This linear framework is efficient and flexible and enables parametric studies of forced responses and energy gains from high-spatio-temporal schlieren data covering long time periods compared to those from scale-resolved simulations.

Results with impulse forcing along a vertical line using both raw and filtered schlieren data to trigger a prototypical shear-layer instability indicate that the perturbation evolves into a wavepacket structure that emits an acoustic disturbance from the splitter-plate trailing edge.
The downstream interaction of this wavepacket with a shock produces a bifurcation of structures traveling at two different speeds, inferred to be fast and slow acoustic waves, in addition to acoustic wavefronts which refract into the subsonic stream.  
In a local receptivity study, the impulse was applied to each pixel in the upper and lower streams to identify the location of the optimal acoustic response.
The highest amplification factor is obtained with perturbations in the upper-stream outer boundary-layer, where acoustic amplification is nine times larger than that of background acoustic disturbances measured in the lower-stream.
On the other hand, instabilities triggered in the lower-stream boundary-layer only yield three-fold acoustic gains.
A harmonic parametric forcing analysis was also conducted across all pixels in the upper and lower streams.
These results revealed multiple tones from the upper stream that were significantly larger than the single tone produced in the lower stream.
Individual responses of these largest harmonic forcings demonstrate spatial structures that adopt the unique disturbance signatures of each forcing location.
While most of the selected conditions trigger the shear-layer instability such that the acoustic response is amplified, one forcing location and frequency was found to be receptive but produced little acoustic radiation.
Results of this type indicate ways by which the DMD-ROM may be used to augment traditional analysis of qualitative schlieren data.

\textbf{Acknowledgments:} The authors are grateful for comments from J.C. Dutton and G.S. Elliott. They also acknowledge support from the Collaborative Center for Aeronautical Sciences and the Office of Naval Research. The first author is also currently supported by an NRC fellowship award at AFRL.

\textbf{Declaration of interest:} The authors declare no conflict of interest.

\appendix
\section{DMD-ROM stabilization}\label{secn:append}
The numerical stability of data-driven reduced order models sometimes poses a challenge, particularly when employed with measurements containing background noise.
The problem manifests in an unbounded solution, as is the case for the filtered schlieren data in this work. 
In the context of DMD, stability characteristics are dependent on the nature of the dynamic operator, $A$, or more precisely, the reduced operator $\tilde{A}$.

To stabilize the DMD-ROM, a POD rank truncation of the first $r$ modes is applied to the derivation of $\tilde{A}$. 
This effectively removes spurious low energy modes which can cause unstable eigenvalues.
The stabilization is implemented by truncating the output of the SVD in Eqn.~\ref{eqn:SVD}, producing $U_r$, $\Sigma_r$, and $V_r$, and resulting in the new dynamic operator
\begin{equation}
     \tilde{A}_r=U_r^* Q_2 V_r \Sigma_r^{-1}.
\end{equation}

\begin{figure}
\centering
\includegraphics[width=1\textwidth]{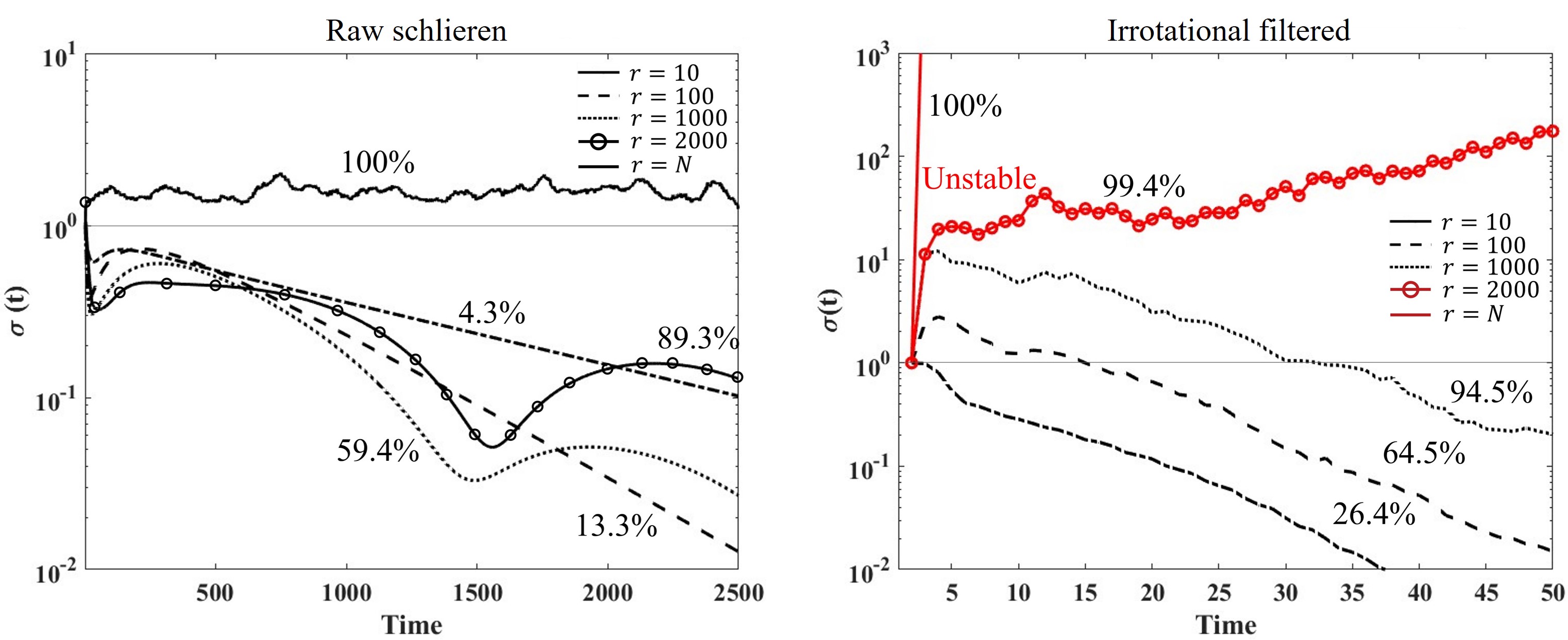}
\caption{Transient gain of the stabilized DMD model for various POD mode rank truncation levels $r$, for the (a) raw and (b) irrotational filtered data.}
\label{fig:r_study}
\end{figure}

The behavior of the truncated ROM is a function of the number of modes adopted, yielding a trade-off between energy retained versus stability.
A brief study of the effect of truncation rank $r$ is presented to determine effects appropriate for the receptivity analysis.
Specifically, the first impulse forcing study of Section~\ref{secn:DMD_ROM} is repeated for values of $r=10, 100, 1{,}000$, and $2{,}000$, with the transient gain used to measure energy behavior.
Results are presented in Fig.~\ref{fig:r_study} for the~(a) raw and~(b) filtered schlieren.
First considering the full ROM ($r=N=2{,}500$), the~(a) raw schlieren is stable while~(b) has unbounded growth after only a few timesteps. 
Figure~\ref{fig:r_study}(b) demonstrates  how the truncation dampens the solution over time, where after a certain value, the decay rate is approximately inversely proportional to $r$. 

Although the gain decay of the truncated ROM drives the solution towards zero over time, the relative trends in amplitude for different forcing parameters such as location and frequency are preserved.
Furthermore, the asymptotic behavior of the solution can be recovered by normalizing the forced snapshots by the transient gain at every instant in time $q'(t)/\sigma (t)$, as shown in Fig.~\ref{fig:impulse_probe}. 





\bibliographystyle{jfm}
\bibliography{current_jfm}

\end{document}